\documentclass{article}

\usepackage{arxiv}

\usepackage[utf8]{inputenc} 
\usepackage[T1]{fontenc} 
\usepackage{hyperref} 
\usepackage{url} 
\usepackage{booktabs} 
\usepackage{amsfonts} 
\usepackage{nicefrac} 
\usepackage{microtype} 
\usepackage{booktabs}
\usepackage{xcolor}
\hypersetup{draft}
\usepackage{graphicx}
\definecolor{darkgreen}{rgb}{0.1, 0.7, 0.3}
\usepackage{listings}
\usepackage{subcaption}

\title{An Introduction to hpxMP: A Modern OpenMP Implementation Leveraging HPX, An Asynchronous Many-Task System}

\author{
 Tianyi Zhang, Shahrzad Shirzad,Patrick Diehl, R. Tohid, Weile Wei, Hartmut Kaiser \\
 Center for Computation and Technology, LSU\\
 \texttt{tzhan18@lsu.edu,sshirz1@lsu.edu,patrickdiehl@lsu.edu,} \\
 \texttt{mraste2@lsu.edu, wwei9@lsu.edu,hkaiser@cct.lsu.edu} \\
}

\begin{document}
\maketitle

\begin{abstract}
Asynchronous Many-task (AMT) runtime systems have gained increasing acceptance in the HPC community due to the performance improvements offered by fine-grained tasking runtime systems. At the same time, C++ standardization efforts are focused on creating higher-level interfaces able to replace OpenMP or OpenACC in modern C++ codes. These higher level functions have been adopted in standards conforming runtime systems such as HPX, giving users the ability to simply utilize fork-join parallelism in their own codes. Despite innovations in runtime systems and standardization efforts users face enormous challenges porting legacy applications. Not only must users port their own codes, but often users rely on highly optimized libraries such as BLAS and LAPACK which use OpenMP for parallization. Current efforts to create smooth migration paths have struggled with these challenges, especially as the threading systems of AMT libraries often compete with the treading system of OpenMP. 

To overcome these issues, our team has developed hpxMP, an implementation of the OpenMP standard, which utilizes the underlying AMT system to schedule and manage tasks. This approach leverages the C++ interfaces exposed by HPX and 
allows users to execute their applications on an AMT system without changing their code.

In this work, we compare hpxMP with Clang's OpenMP library with four linear algebra benchmarks of the Blaze C++ library. While hpxMP is often not able to reach the same performance, we demonstrate viability for providing a smooth migration for applications but have to be extended to benefit from a more general task based programming model.
\end{abstract}


\twocolumn 

\section{Introduction}
The Open Multi-Processing (OpenMP)~\cite{openmp_spec} standard is widely used for shared memory multiprocessing and is often coupled with the Message Passing Interface (MPI) as \textit{MPI+X} \cite{10.1109/MC.2016.232} for distributed programming. Here, MPI is used for the inter-node communication and \textit{X}, in this case, OpenMP, for the intra-node parallelism. Nowadays, Asynchronous Many Task (AMT) run time systems are emerging as a new parallel programming paradigm. These systems are able to take advantage of fine grained tasks to better distribute work across a machine. The C++ standard library for concurrency and parallelism (HPX)~\cite{heller2017hpx} is one example of an AMT runtime system. The HPX API conforms to the concurrency abstractions introduced by the C++ 11 standard~\cite{cxx11_standard} and to the parallel algorithms introduced by the C++ 17 standard~\cite{cxx17_standard}. These algorithms are similar to the concepts exposed by OpenMP, \emph{e.g.} \lstinline|#pragma omp parallel for|. 

AMT runtime systems are becoming increasing used for HPC applications as they have shown superior scalability and parallel efficiency for certain classes of applications (see ~\cite{heller_gb}). At the same time, the C++ standardization efforts currently focus on creating higher-level interfaces usable to replace OpenMP (and other \lstinline|#pragma|-based parallelization solutions like OpenACC) for modern C++ codes. This effort is driven by the lack of integration of \lstinline|#pragma| based solutions into the C++ language, especially the language's type system.

Both trends call for a migration path which will allow existing applications that directly or indirectly use OpenMP to port potions of the code an AMT paradigm. This is especially critical for applications which use highly optimized OpenMP libraries where it is not feasible to re-implement all the provided functionalities into a new paradigm. Examples of these libraries are linear algebra libraries~\cite{eigenweb,rupp2016viennacl,sanderson2016armadillo,iglberger2012high,galassi2002gnu,wang2013augem,laug,blackford2002updated}, such as the Intel math kernel library or the Eigen library.

For these reasons, it is beneficial to combine both technologies, \textit{AMT+OpenMP}, where the distributed communication is handled by the AMT runtime system and the intra-node parallelism is handled by OpenMP or even combine OpenMP and the parallel algorithms on a shared memory system. Currently, these two scenarios are not possible, since the light-weighted thread implementations usually present in AMTs interferes with the system threads utilized by the available OpenMP implementations. 

To overcome this issue, hpxMP, an implementation of the OpenMP standard~\cite{openmp_spec} that utilizes HPX's light-weight threads is presented in this paper. The hpxMP library is compatible with the clang and gcc compiler and replaces their shared library implementations of OpenMP. hpxMP implements all of the OpenMP runtime functionalities using HPX's lightweight threads instead of system threads. 

Blaze, an open source, high performance C++ math library, \cite{iglberger2012high} is selected as an example library to validate our implementation. Blazemark, the benchmark suite available with Blaze, is used to run some common benchmarks. The measured results are compared against the same benchmarks run on top of the compiler-supplied OpenMP runtime. This paper focuses on the implementation details of hpxMP as a proof of concept implementing OpenMP with an AMT runtime system. We use HPX as an exemplary AMT system that already exposes all the required functionalities. 

The paper is structured as follows: Section~\ref{sec:rel:work} emphasizes the related work. Section~\ref{sec:hpx} provides a brief introduction to HPX's concepts and Section~\ref{sec:hpxMP} a brief introduction to OpenMP's concepts utilized in the implementation in Section~\ref{sec:impl}. The benchmarks comparing hpxMP with clang's OpenMP implementation are shown in Section~\ref{sec:bench}. Finally, we draw our conclusions in Section~\ref{sec:con}.

\section{Related Work}
\label{sec:rel:work}
Exploiting parallelism on multi-core processors with shared memory has been
extensively studied and many solutions have been implemented. The POSIX
Threads~\cite{pthreads} execution model enables fine grain parallelism
independent of any language. At higher levels, there are also library solutions
like Intel's Threading Building Blocks (TBB)~\cite{inteltbb} and Microsoft's
Parallel Pattern Library (PPL)~\cite{microsoftppl}. TBB is a C++ template
library for task parallelism while PPL provides features like task parallelism,
as well as parallel algorithms and containers with an imperative programming model.
There are also several language solutions.
Chapel~\cite{chamberlain07parallelprogrammability} is a parallel programming
language with parallel data and task abstractions. The Cilk family of
languages~\cite{cilk++} are general-purpose programming languages which target
multi-thread parallelism by extending C/C++ with parallel loop constructs and
a fork-join model. Kokkos~\cite{CarterEdwards20143202} is a package which exposes multiple
parallel programming models such as CUDA and pthreads through a common C++
interface. Open Multi-Processing (OpenMP)~\cite{openmp1} is a widely accepted standard used by
application and library developers. OpenMP exposes fork-join model through
compiler directives and supports tasks.

The OpenMP $3$.$0$ standard\footnote{https://www.openmp.org/wp-content/uploads/spec30.pdf} introduced the concept of task-based programming. The OpenMP $3$.$1$ standard\footnote{https://www.openmp.org/wp-content/uploads/OpenMP3.1.pdf} added task optimization within the tasking model. The OpenMP $4$.$0$ standard\footnote{https://www.openmp.org/wp-content/uploads/OpenMP4.0.0.pdff} offers user a more graceful and efficient way to handle task synchronization by introducing depend tasks and task group. The OpenMP $4$.$5$ standard\footnote{https://www.openmp.org/wp-content/uploads/openmp-4.5.pdf} was released with its support for a new task-loop construct, which is providing a way to separate loops into tasks. The most recent, the OpenMP $5$.$0$ standard\footnote{https://www.openmp.org/wp-content/uploads/OpenMP-API-Specification-5.0.pdf} supports detached tasks.

There have also been efforts to integrate multi-thread parallelism with
distributed programming models. Charm++ has integrated OpenMP into its
programming model to improve load balance~\cite{ppl_link}. However, most of the
research in this area has focused on MPI+X~\cite{10.1109/MC.2016.232,
barrett2015toward} model.

\section{C++ Standard Library for Concurrency and Parallelism (HPX)}
\label{sec:hpx}
This section briefly describes the features of the C++ Standard Library for Concurrency and Parallelism (HPX)~\cite{heller2017hpx} which are utilized in the implementation of hpxMP in the Section~\ref{sec:impl}. HPX facilitates distributed parallel applications of any scale and uses fine-grain multi-threading and asynchronous communications~\cite{khatami2016massively}. HPX exposes an API that strictly adheres the current ISO C++ standards~\cite{heller2017hpx}. This approach to standardization encourages programmers to write code that is high portability in heterogeneous systems~\cite{copik2017using}.

HPX is highly interoperable in distributed parallel applications, such that, it can be used on inter-node communication setting of a single machine as well as intra-node parallelization scenario of hundreds of thousands of nodes~\cite{wagle2018methodology}. The \textit{future} functionality implemented in HPX permits threads to continually finish their computation without waiting for their previous steps to be completed which can achieve a maximum possible level of parallelization in time and space~\cite{khatami2017redesigning}.
\subsection{HPX Threads}
The HPX light-weight threading system provides user level threads, which enables fast context switching~\cite{biddiscombezero}. With lower overheads per thread, programs are able to create and schedule a large number of tasks with little penalty 
~\cite{wagle2018methodology}. 
The advantage of this threading system combined with the \textit{future} functionality in HPX facilitates auto-parallelization in a highly efficient fashion as such combination allows the direct expression of the generated dependency graph as an execution tree generated at runtime~\cite{grubel2015performance}.
\subsection{HPX Thread Scheduling and Policies}
The adaptive thread scheduling system employed in HPX improves performance of parallel applications~\cite{biddiscombezero}. The built-in scheduling policies enable optimal task scheduling for a given application and/or algorithm~\cite{heller2017hpx}. The programmers can code efficiently as they can focus on algorithms or application development itself instead of manually scheduling CPU resources. Also, the built-in scheduling policies allow users to provide their own scheduling policies if they require more specific control on application-level.

The HPX runtime now supports eight different thread scheduling policies: \textit{priority local scheduling (default option)}: this policy creates one queue per OS thread. The OS threads remove waiting tasks from the queue and start task execution accordingly. The number of high priority queues equal to the number of OS threads. \textit{Static priority scheduling}: the static priority scheduling policy maintains one queue per OS thread from which each OS thread places its tasks. Round Robin model is used in this policy. \textit{Local scheduling}: this policy maintains one queue per OS threads from which each OS thread removes waiting tasks from the queue and start task execution accordingly. \textit{Global scheduling}: this policy maintains one shared queue from which all OS threads pull waiting tasks. \textit{ABP scheduling}: this policy maintains a double ended lock-free queue per OS thread. Threads are inserted on the top of the queue and are stolen from the bottom of the queue during the work stealing. \textit{Hierarchy scheduling policy}: this policy constructs a tree of task items, and each OS thread traverses through the tree to obtain new task item. \textit{Periodic priority scheduling policy}: this policy arranges one queue of task items per OS thread, a couple of high priority queues and one low priority queue. 

These policies can be categorized into three types: \textit{thread local}: the thread local is currently set as the default scheduling option. This policy schedules one queue for each OS core, and the queues with high priority will be scheduled before any other works with lower priority; \textit{static}: the static scheduling policy follows round robin principle and the thread stealing is not allowed in this policy; \textit{hierarchical}: the hierarchical scheduling policy utilizes a tree structure of run queues. The OS threads need to traverse the tree to new task items.

\section{Integration of HPX in OpenMP Applications}
\label{sec:hpxMP}
\begin{figure}[tb]
	\centering
	\includegraphics[width=\linewidth,trim={0 5.5cm 0 3.0cm},clip]{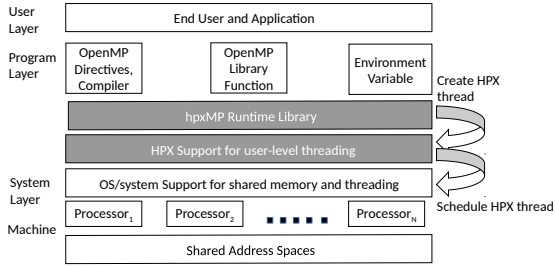}
	\caption{Layers of an hpxMP application. Applications are compiled with
	OpenMP flags and linked against the hpxMP runtime library where HPX threads
	perform the computation in parallel. Gray layers are implemented
	in hpxMP. This figure is adapted from \cite{openmpintro}.}.
	\label{fig:hpxmp_struct}
\end{figure}

This section describes the integration of HPX to the OpenMP
4.0\footnote{https://www.openmp.org/wp-content/uploads/OpenMP4.0.0.pdf} specification. Figure~\ref{fig:hpxmp_struct} illustrates how hpxMP fits in an OpenMP application. A user-defined application with OpenMP directives, library
functions, and environment variable can be compiled with any compiler that
supports OpenMP. The hpxMP shared library adds an additional layer, marked in
gray, which carries out the parallel computation. Instead of calling the OpenMP
functions and running them on OpenMP threads, the equivalent hpxMP functions are
called redirecting the program to the corresponding functionality in HPX.
HPX employs light-weight HPX threads following thread scheduling policies for
parallel computing. For more details see Section~\ref{sec:hpx}.

Table~\ref{tab:directives} provides the list of the directives that are
currently supported by hpxMP. A list of runtime library functions implemented in
hpxMP runtime library is available in Table~\ref{tab:library}. In the next
section, the detailed implementation of these directives is discussed.

\begin{table*}[tb]
\centering
	\caption{Directives implemented in the program layer of hpxMP, see Figure~\ref{fig:hpxmp_struct}. The corresponding functions are the main part of hpxMP runtime library.}
	\label{tab:directives}
	\begin{tabular}{ccl}
		\toprule
		Pragmas Implemented in hpxMP 	\\
		\midrule
\#pragma omp atomic & \#pragma omp barrier\\
\#pragma omp critical & \#pragma omp for	\\
\#pragma omp master & \#pragma omp ordered \\
\#pragma omp parallel & \#pragma omp section \\
\#pragma omp single & \#pragma omp task depend	\\
		\bottomrule
	\end{tabular}
\end{table*}

\begin{table*}[tb]
\centering
	\caption{Runtime library functions in hpxMP's program layer, see Figure ~\ref{fig:hpxmp_struct}. The following functional APIs are provided to users. }
	\label{tab:library}
	\begin{tabular}{ccl}
		\toprule
		Runtime Library Functions in hpxMP 	\\
		\midrule
		\lstinline|omp_get_dynamic| & \lstinline|omp_get_max_threads| \\
		\lstinline|omp_get_num_procs| &
		\lstinline|omp_get_num_threads|	\\
		\lstinline|omp_get_thread_num|&
		\lstinline|omp_get_wtick|	\\
		\lstinline|omp_get_wtime|&
		\lstinline|omp_in_parallel|	\\
		\lstinline|omp_init_lock|&
		\lstinline|omp_init_nest_lock|	\\
		\lstinline|omp_set_dynamic|&
		\lstinline|omp_set_lock|	\\
		\lstinline|omp_set_nest_lock|	&
		\lstinline|omp_set_num_threads|	\\
		\lstinline|omp_test_lock|	&
		\lstinline|omp_test_nest_lock|	\\
		\lstinline|omp_unset_lock|	&
		\lstinline|omp_unset_nest_lock|	\\
		\bottomrule
	\end{tabular}
\end{table*}

\section{Implementation of hpxMP}
\label{sec:impl}
\begin{table}[tb]
\centering
	\caption{OMPT callbacks implemented in hpxMP runtime library, see Figure~\ref{fig:hpxmp_struct}. First party performance analysis toolkit for users to develop higher level performance analysis policy.}
	\label{tab:ompt}
	\begin{tabular}{ccl}
		\toprule
		OMPT callbacks	\\
		\midrule
		ompt\_callback\_thread\_begin	\\
		ompt\_callback\_thread\_end \\
		ompt\_callback\_parallel\_begin	\\
		ompt\_callback\_parallel\_end\\
		ompt\_callback\_task\_create	\\
		ompt\_callback\_task\_schedule	\\
		ompt\_callback\_implicit\_task	\\
		\bottomrule
	\end{tabular}
\end{table}

This section provides an overview of prominent OpenMP and OpenMP Performance
Toolkit (OMPT) functionalities and elaborates on implementations of these
functionalities in hpxMP.

The fundamental directives described in the OpenMP specification~\cite{openmptr} are
shown in Listing~\ref{lst:pragma}. It is important to note that implementation
of OpenMP directives may differ based on the compiler. hpxMP is mapped onto 
LLVM-Clang as specified by LLVM OpenMP Runtime
Library{\footnote{https://openmp.llvm.org/Reference.pdf}}. However, hpxMP also
supports GCC{\footnote{https://gcc.gnu.org/onlinedocs/libgomp/}} entry by mapping
the function calls generated by the GCC compiler on to the Clang entry.

\begin{lstlisting}[caption=Fundamental OpenMP directives,label={lst:pragma}]
#pragma omp parallel
#pragma omp for
#pragma omp task
\end{lstlisting}

\subsection{Parallel Construct}
Parallel construct initiates the parallel execution of the portion of the code
annotated by the parallel directive. The directive \lstinline|#pragma omp parallel| with its associated structured block is treated as a function call to
\lstinline|__kmpc_fork_call|, which preprocesses the arguments passed by the
compiler and calls the function named \lstinline|fork|, implemented in hpxMP
runtime, see Figure~\ref{fig:hpxmp_struct}. The implementation of
\lstinline|__kmpc_fork_call| is shown in Listing~\ref{lst:parallel}. HPX
threads, as many as requested by the user, are created during the fork call which explicitly registers HPX threads, see
Listing~\ref{lst:fork}. Each HPX thread follows HPX scheduling policies, see
Section~\ref{sec:hpx}, performing its own work under the structured parallel block.

\begin{lstlisting}[caption=Implementation of \_\_kmpc\_fork\_call in hpxMP,label={lst:parallel},float=*ptb,numbers=left]
void __kmpc_fork_call(ident_t *loc, kmp_int32 argc, kmpc_micro microtask, ...) {
	vector<void*> argv(argc);
	va_list ap;
	va_start( ap, microtask );
	for( int i = 0; i < argc; i++ ){
		argv[i] = va_arg( ap, void * );
	}
	va_end( ap );
	void ** args = argv.data();
	hpx_backend->fork(__kmp_invoke_microtask, microtask, argc, args);
}
\end{lstlisting}

\begin{lstlisting}[caption=Implementation of hpx\_runtime::fork in hpxMP,label={lst:fork},float=*ptb,numbers=left]
for( int i = 0; i < parent->threads_requested; i++ ) {
	hpx::applier::register_thread_nullary(
		std::bind( &thread_setup, kmp_invoke, thread_func, argc, argv, i, &team, parent,
		boost::ref(barrier_mtx), boost::ref(cond), boost::ref(running_threads) ),
		"omp_implicit_task", hpx::threads::pending,
		true, hpx::threads::thread_priority_low, i );
	}
\end{lstlisting}

\subsection{Loop Scheduling Construct}
Another common OpenMP construct is the loop construct which runs several
iterations of a \lstinline|for| loop in parallel where each instance runs on a
different thread in the team. A team is defined as a set of one or more threads in
the execution of a parallel region~\cite{openmptr}. The loops are divided into
chunks, and the scheduler determines how such chunks are distributed across the
threads in the team. The default schedule type is \lstinline|static| where the
chunk size is determined by the threads and number of loops. Each thread gets
approximately the same amount of loops and the structured block is executed in
parallel. 

For the static schedule, the directive \lstinline|#pragma omp for| with its
associated structured for loop block invoke the following sequence of function
calls: \lstinline|__kmpc_for_static_init|, \lstinline|__kmpc_dispatch_next| and
\lstinline|__kmpc_dispatch_fini|. Chunks are distributed among threads in a
round-robin fashion, see Listing~\ref{lst:loop}.
\begin{lstlisting}[caption=Implementation of \_\_kmpc\_for\_static\_init in hpxMP,label={lst:loop},float=*ptb,numbers=left]
void __kmpc_for_static_init( ident_t *loc, int32_t gtid, int32_t schedtype, 
		int32_t *p_last_iter, int64_t *p_lower, int64_t *p_upper, 
		int64_t *p_stride, int64_t incr, int64_t chunk ) {
	//code to determine each thread's lower and upper bound (*p_lower, *p_upper) 
	//with the given thread id, schedule type and stride.
	}
\end{lstlisting}

\subsection{Task Construct}
Task Construct creates explicit tasks in hpxMP. When a thread sees this
construct, a new HPX thread is created and scheduled based on HPX thread
scheduling policies, see Section~\ref{sec:hpx}. 

The directive \lstinline|#pragma omp task| and its associated structured block
initiate a series of function calls:
\lstinline|__kmpc_omp_task_alloc|, \lstinline|__kmpc_fork_call|, see Listing
~\ref{lst:task}. 

A task object is allocated, initialized, and returned to the
compiler by the task allocation function \lstinline|__kmpc_omp_task_alloc|. A normal priority HPX thread is then created by the compiler generated function call \lstinline|__kmpc_omp_task| and ready to execute the task allocated by prior functions.

\begin{lstlisting}[caption=Implementation of task scheduling in hpxMP,label={lst:task},float=*ptb,numbers=left]
kmp_task_t* __kmpc_omp_task_alloc( ident_t *loc_ref, kmp_int32 gtid, kmp_int32 flags,
			size_t sizeof_kmp_task_t, size_t sizeof_shareds,
			kmp_routine_entry_t task_entry ){
	int task_size = sizeof_kmp_task_t + (-sizeof_kmp_task_t%8);
	kmp_task_t *task = (kmp_task_t*)new char[task_size + sizeof_shareds]; 
	task->routine = task_entry;
	return task;
}
int __kmpc_omp_task( ident_t *loc_ref, kmp_int32 gtid, kmp_task_t * new_task){
		//Create a normal priority HPX thread with the allocated task as argument.
		hpx::applier::register_thread_nullary(.....)
		return 1;
}
\end{lstlisting}

\subsection{OpenMP Performance Toolkit}
OpenMP Performance Toolkit (OMPT) is an application programming interface (API) for first-party performance
tools. It is integrated into the hpxMP runtime system and enables users to construct
powerful and efficient custom performance tools. The implemented callback
functions, see Table ~\ref{tab:ompt}, make it possible for users to track the
behavior of threads, parallel regions, and tasks. 

The Implementation of the parallel begin callback is shown in
Listing~\ref{lst:callback_parall}. This piece of code calls the user-defined
callbacks and is hooked into the hpxMP runtime before the parallel region
actually begins. 
\begin{lstlisting}[caption=Implementation of thread callbacks in hpxMP,label={lst:callback_parall},float=*ptb,numbers=left]
if (ompt_enabled.enabled) {
	if (ompt_enabled.ompt_callback_parallel_begin) {
		ompt_callbacks.ompt_callback(ompt_callback_parallel_begin)(
			NULL, NULL,&team.parallel_data, team_size,
			__builtin_return_address(0));
	}
}
#endif
\end{lstlisting}

\subsection{GCC Support}
LLVM OpenMP Runtime
Library{\footnote{https://openmp.llvm.org/Reference.pdf}} provides the gcc compatibility shims. In order to achieve the GCC support in hpxMP, we exposes similar shims to map GCC generated entries to
Clang. These mapping functions preprocess the arguments provided by the compiler
and pass them directly to the hpxMP or call Clang supported entries. Therefore,
programs compiled with GCC or Clang are supported by hpxMP.

\begin{lstlisting}[caption=Implementation of gcc entry fork call in hpxMP,label={lst:fork_gcc},float=*ptb,numbers=left]
void
xexpand(KMP_API_NAME_GOMP_PARALLEL)(void (*task)(void *), void *data, unsigned num_threads, unsigned int flags) {
	omp_task_data * my_data = hpx_backend->get_task_data();
	my_data->set_threads_requested(num_threads);
	__kmp_GOMP_fork_call(task,(microtask_t )__kmp_GOMP_microtask_wrapper, 2, task, data);
}
\end{lstlisting}

\subsection{Start HPX back end}
HPX must be initialized before hpxMP can start execution. The HPX initialization can start both externally or internally. If HPX is started externally (by applications), hpxMP will initialize HPX internally before scheduling any work. The function designed to start hpx back-end properly is written in each function calls generated by the compiler make sure HPX is properly started before we call any \lstinline|#pragma omp| related functions, see Listing~\ref{lst:hpx_start}.

\begin{lstlisting}[caption=Implementation of starting HPX in hpxMP,label={lst:hpx_start},float=*ptb,numbers=left]
 hpx::start(f, desc_cmdline, argc, argv, cfg, std::bind(&wait_for_startup, boost::ref(startup_mtx), boost::ref(cond), boost::ref(running)));
\end{lstlisting}

\section{Benchmarks}
\label{sec:bench}
In this paper, four benchmarks are used to compare the performance between Clang's implementation of OpenMP and our implementation of hpxMP, which are daxpy, dense vector addition, dense matrix addition, and dense matrix multiplication. We tested these benchmarks are tested on Marvin, a node in the Center of Computation and Technology (CCT)'s Rostam cluster at Louisiana State University. The hardware properties of Marvin are shown in Table~\ref{tab:system} and the libraries and compiler used to build hpxMP and its dependencies can be found in Table~\ref{tab:software}. 

The benchmark suite of Blaze{\footnote{https://bitbucket.org/blaze-lib/blaze/wiki/Benchmarks}} is used to analyze the performance. For each benchmark, a heat-map demonstrates the ratio, $r$, of the Mega Floating Point Operations Per Second (MFLOP/s) between hpxMP and OpenMP. To make the heat map plots easier to analyze, only a portion of the larger input sizes $n$ is visualized. For the overall overview, three scaling graphs with thread number $4$, $8$, and $16$ are plotted associated with each benchmark, showing the relation between MFLOP/s and size $n$. The size of the vectors and matrix in the benchmarks increases arithmetically from $1$ to $10$ million. We picked these three thread numbers as an example since the behavior looks similar for all other candidates.
Blaze uses a set of thresholds for different operations to be executed in parallel. For each of the following benchmarks if the number of elements in the vector or matrix (depending on the benchmark) is smaller than the specified threshold for that operation, it would be executed single-threaded.

\begin{table*}[tb]
\centering
	\caption{System configuration of the marvin node. All benchmarks are run on this node.}
	\label{tab:system}
	\begin{tabular}{ccl}
		\toprule
		Category & Property	\\
		\midrule
		Server Name & Rostam	\\
		CPU & 2 x Intel(R) Xeon(R) CPU E5-2450 0 @ 2.10GHz \\
		RAM & 48 GB	\\
		Number of Cores & 16\\
		\bottomrule
	\end{tabular}
\end{table*}

\begin{table}[tb]
\centering
	\caption{Overview of the compilers, software, and operating system used to build hpxMP, HPX, Blaze and its dependencies.}
	\label{tab:software}
	\begin{tabular}{ccl}
		\toprule
		Category & Property	\\
		\midrule
		OS & CentOS Linux release 7.6.1810 (Core) \\
		Kernel & 3.10.0-957.1.3.el7.x86\_64 \\
		Compiler & clang 6.0.1 \\
		gperftools & 2.7	\\
		boost & 1.68.0	\\
		OpenMP & 3.1	\\
		HPX\footnote{\url{https://github.com/STEllAR-GROUP/hpx}} & 140b878	\\
		Blaze\footnote{\url{https://bitbucket.org/blaze-lib/blaze}} & 3.4 \\
		\bottomrule
	\end{tabular}
\end{table}

\subsection{Dense Vector Addition}
Dense Vector Addition(dvecdvecadd) is a benchmark that adds two dense vectors $a$ and $b$ and stores the result in vector $c$, where $a,b\in \mathbb{R}^n$. The addition operation is $c[i] = a[i]+b[i]$. The parallelization threshold for daxpy benchmark is set to 38000. So we expect to see the effect of parallelization only when the vector size gets greater than or equal to $38000$.

The ratio of performance $r$ is shown in Figure\ref{fig:ratio,dense}. For small vectors $\leq 103,258$ hpxMP scales less than OpenMP especially when the thread number is large, but gets closer OpenMP as the vector size is increasing. Compared to OpenMP, the best performance of hpxMP is achieved between vector size $431,318$ to $2,180,065$ and the threads number between $1$ to $7$. Except for some outliers, hpxMP is between $0$\% and $30$\% slower than the optimized OpenMP version for larger vector sizes . 
The scaling plots are shown in Figure\ref{fig:scale,dvecdvecadd}. For all different number of threads, we see that both implementations behave similar until the parallelization starts. For vector sizes between $10^5$ and $10^6$ hpxMP is slower. For larger input sizes the implementations are comparable again. 

\begin{figure}[h]
	\centering
	\includegraphics[width=\linewidth]{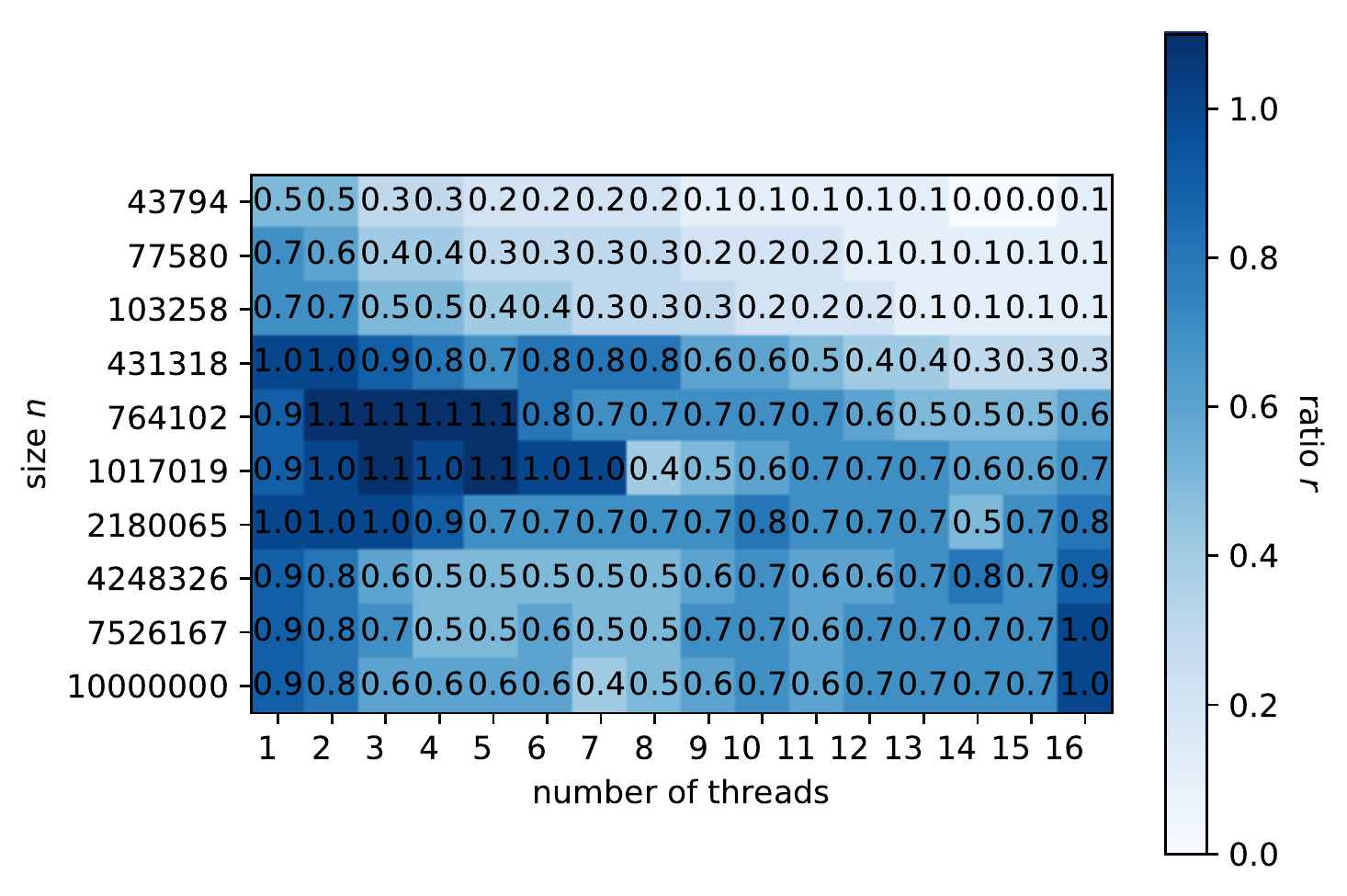}
	\caption{Performance Ratio using dvecdvecadd Benchmark (hpxMP/OpenMP)}
	\label{fig:ratio,dense}
\end{figure}

\subsection{Daxpy}

Daxpy is a benchmark to multiply a number $\beta$ with a dense vector $a$, add the result with a dense vector $b$, and store the result in same vector $b$, where $\beta\in \mathbb{R}$, and $a,b\in \mathbb{R}^n$. The operation used for this benchmark is $b[i] = b[i]+3.0*a[i]$. Same as dvecdvecadd benchmark, the parallelization threshold for daxpy benchmark is set to $38,000$. So we expect to see the effect of parallelization only when the vector size gets $\geq 38,000$.

The ratio of performance $r$ is shown in Figure\ref{fig:ratio,daxpy}. For small vectors $\leq 103,258$ hpxMP scales less than OpenMP especially when the thread size is large but gets closer OpenMP as the vector size is increasing. Compared to OpenMP, the best performance of hpxMP is achieved between vector size $431,318$ to $1,017,019$ and the threads number between $1$ to $8$. Except for some exceptions, hpxMP is between $0$\% and $40$\% slower than the optimized OpenMP version for larger vector sizes
The scaling plot is shown in Figure\ref{fig:scale,daxpy} shows that for all different number of threads, we see that both implementations behave similarly until the parallelization starts. For vector sizes between $10^5$ and $10^6$ hpxMP is slower but for larger input sizes the implementations are comparable again. 

\begin{figure}[h]
	\centering
	\includegraphics[width=\linewidth]{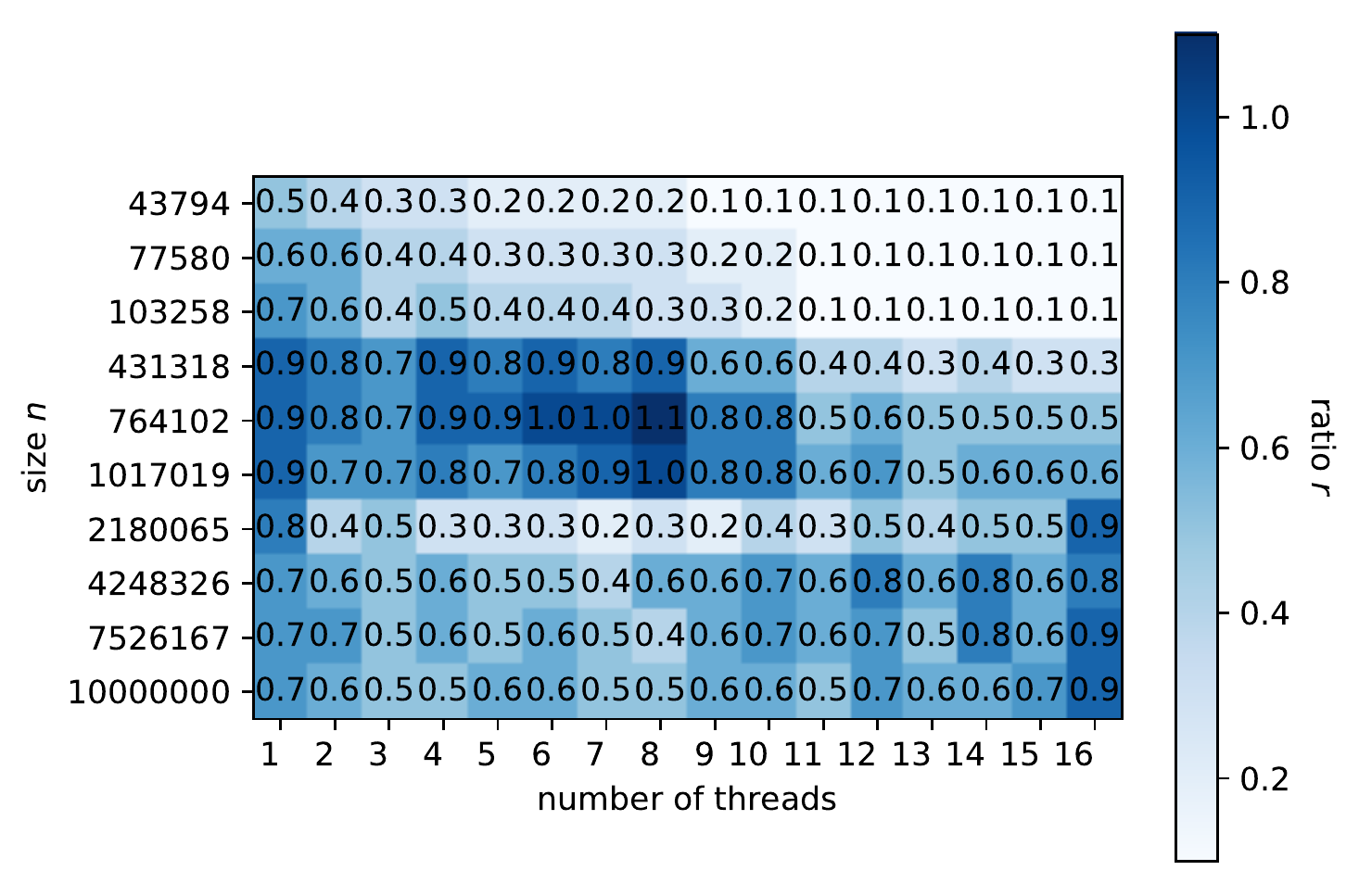}
	\caption{Performance Ratio using daxpy Benchmark (hpxMP/OpenMP)}
	\label{fig:ratio,daxpy}
\end{figure}

\subsection{Dense Matrix Addition}
Dense Matrix Addition(dmatdmatadd) is a benchmark to add two dense matrix$A$ and $B$ and stores the result in matrix $C$, where $A,B\in \mathbb{R}^{n\times n}$. The matrix addition operation is $C[i,j]=A[i,j]+B[i,j]$. The ratio of performance $r$ is shown in Figure\ref{fig:ratio,dmatdmatadd}. The scaling plot is shown in Figure\ref{fig:ratio,dmatdmatadd}. For the dmatdmatadd benchmark, the parllelization threshold set by Blaze is $36,100$. Whenever the target matrix has more than or equal to $36,100$ elements (corresponding to matrix size $190$ by $190$), this operation is executed in parallel. 

Figure\ref{fig:ratio,dmatdmatadd} shows that OpenMP performs better especially when the matrix size is small and the number of thread is large. For a larger number of threads, hpxMP gets closer to OpenMP for larger matrix sizes. Except for some exceptions, hpxMP is between $0$\% and $40$\% slower than the optimized OpenMP version.
Figure\ref{fig:scale,dmatdmatadd} shows that for all different number of threads, we see that both implementations behave similar until the parallelization starts. For matrix sizes between $230$ and $455$ hpxMP is slower but the implementations are comparable again as the input size is increasing. 

\begin{figure}[h]
	\centering
	\includegraphics[width=\linewidth]{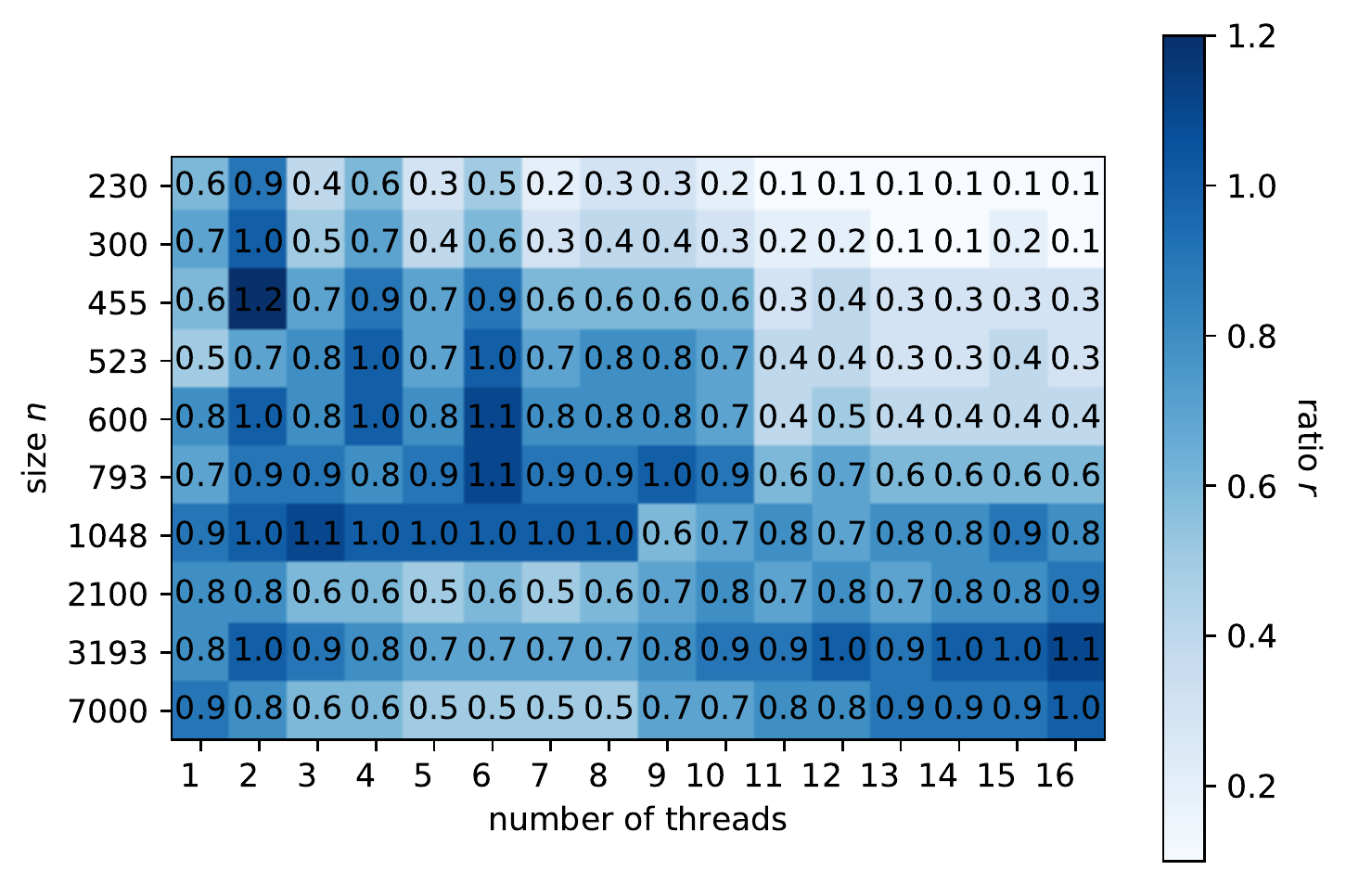}
	\caption{Performance Ratio using dmatdmatadd Benchmark (hpxMP/OpenMP)}
	\label{fig:ratio,dmatdmatadd}
\end{figure}

\subsection{Dense Matrix Multiplication}
Dense Matrix Multiplication(dmatdmatmult) is a benchmark that multiplies two dense matrix $A$ and $B$ and stores the result in matrix $C$, where $A,B\in \mathbb{R}^{n\times n}$. The matrix addition operation is $C=A*B$. The ratio of performance $r$ is shown in Fig.\ref{fig:ratio,dmatdmatmult}. The scaling plot is shown in Fig.\ref{fig:scale,dmatdmatmult}. For the dmatdmatmult benchmark, the parallelization threshold set by Blaze is 3,025. Whenever the target matrix has more than or equal to $36,100$ elements (corresponding to matrix size $55$ by $55$), this operation is executed in parallel. 

Fig.\ref{fig:ratio,dmatdmatmult} shows that OpenMP outperforms hpxMP only when the matrix size is between $230$ and $300$ and the number of thread is between $12$ to $16$. hpxMP gets as fast as OpenMP for other vector sizes.
Fig.\ref{fig:scale,dmatdmatmult} shows that for all different number of threads, we see that both implementations behave similar until the parallelization starts. For matrix sizes between $74$ and $113$ hpxMP is slower. For larger input sizes, the implementations are comparable again. 

\begin{figure}[h]
	\centering
	\includegraphics[width=\linewidth]{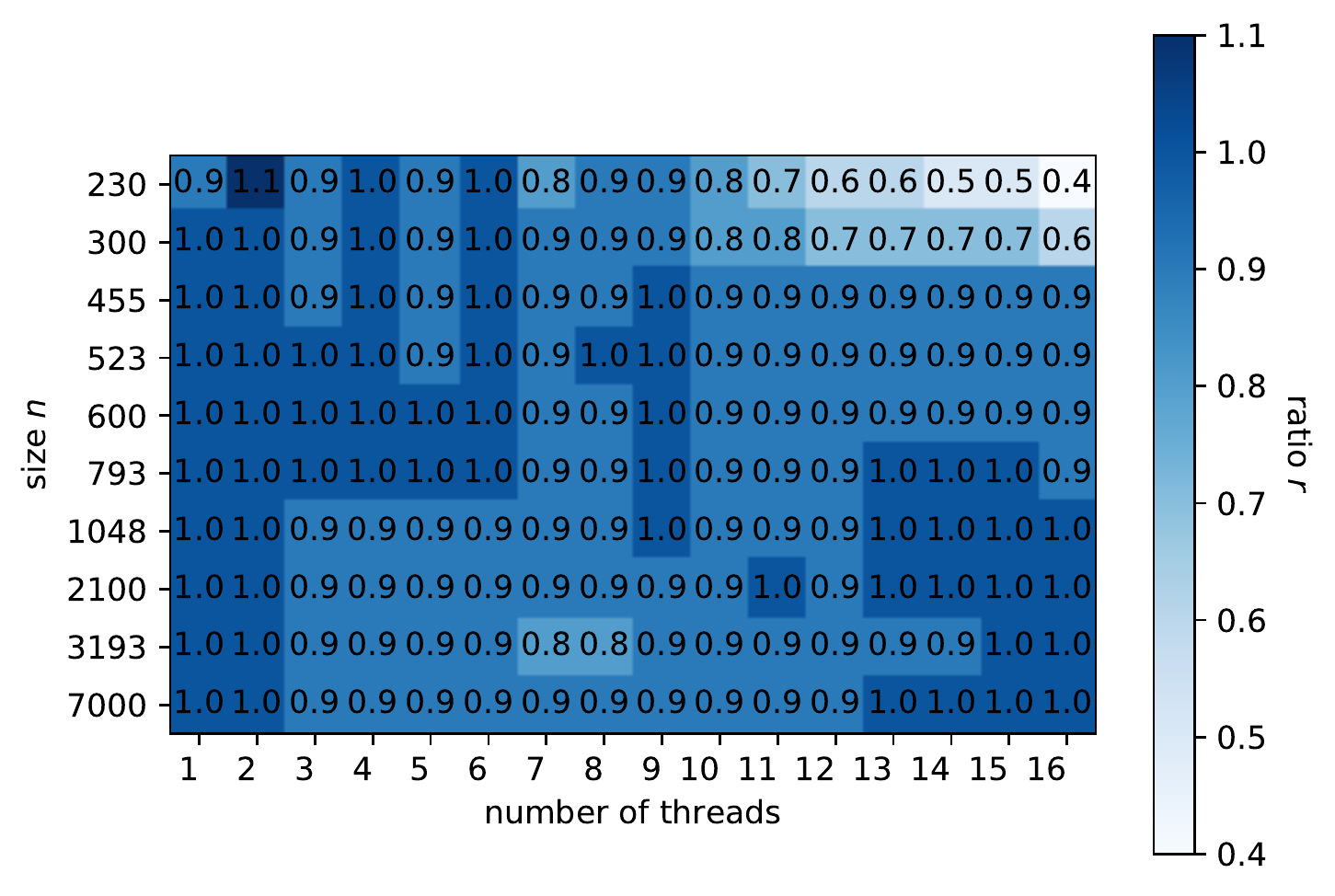}
	\caption{Performance Ratio using dmatdmatmult Benchmark (hpxMP/OpenMP)}
	\label{fig:ratio,dmatdmatmult}
\end{figure}

\begin{figure*}[p]
	\centering
	\begin{subfigure}[b]{0.3\textwidth}
		\includegraphics[width=\linewidth]{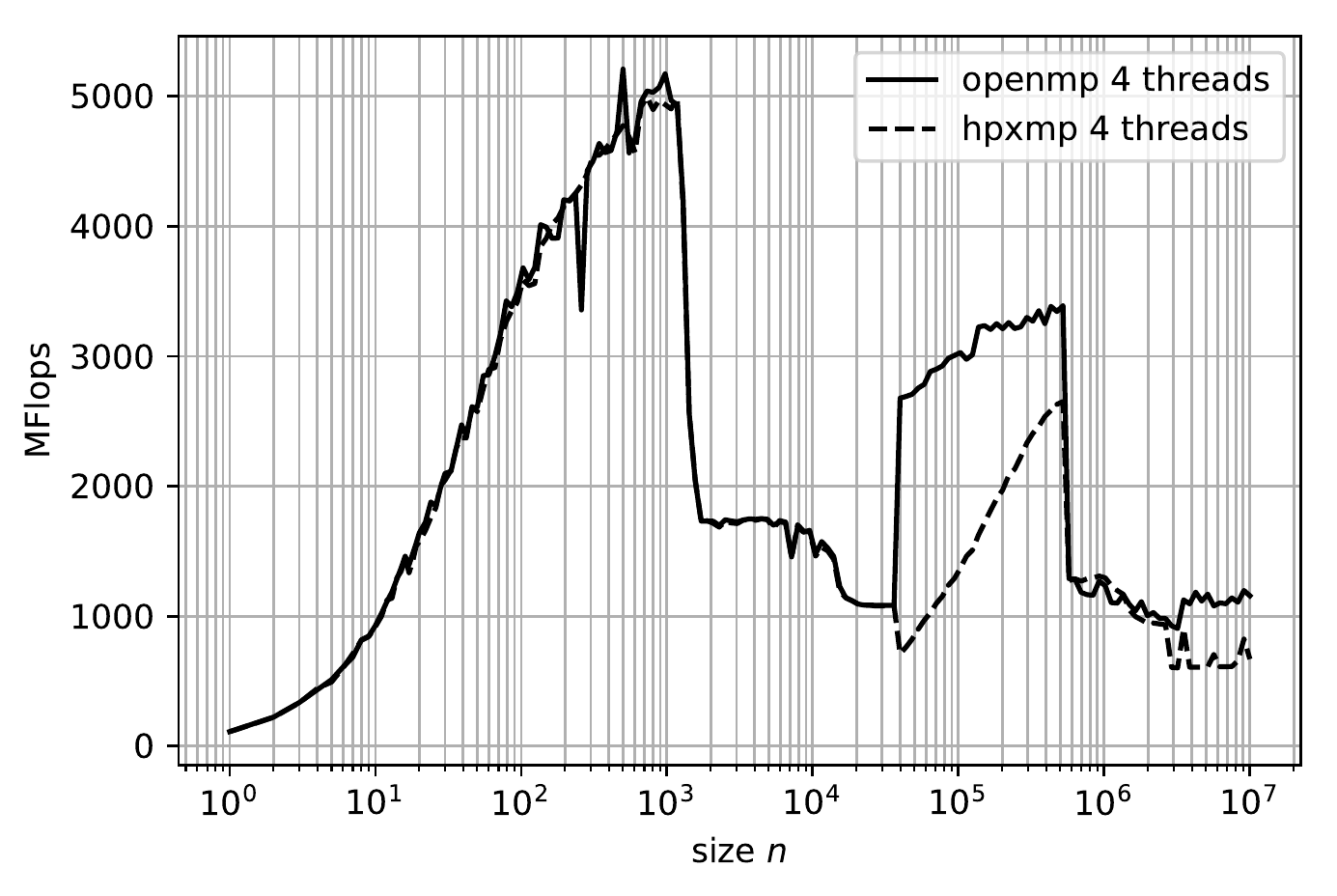}
		\caption{}
		\label{fig:ratio,dvecdvecadd_4}
	\end{subfigure}
	\begin{subfigure}[b]{0.3\textwidth}
		\includegraphics[width=\linewidth]{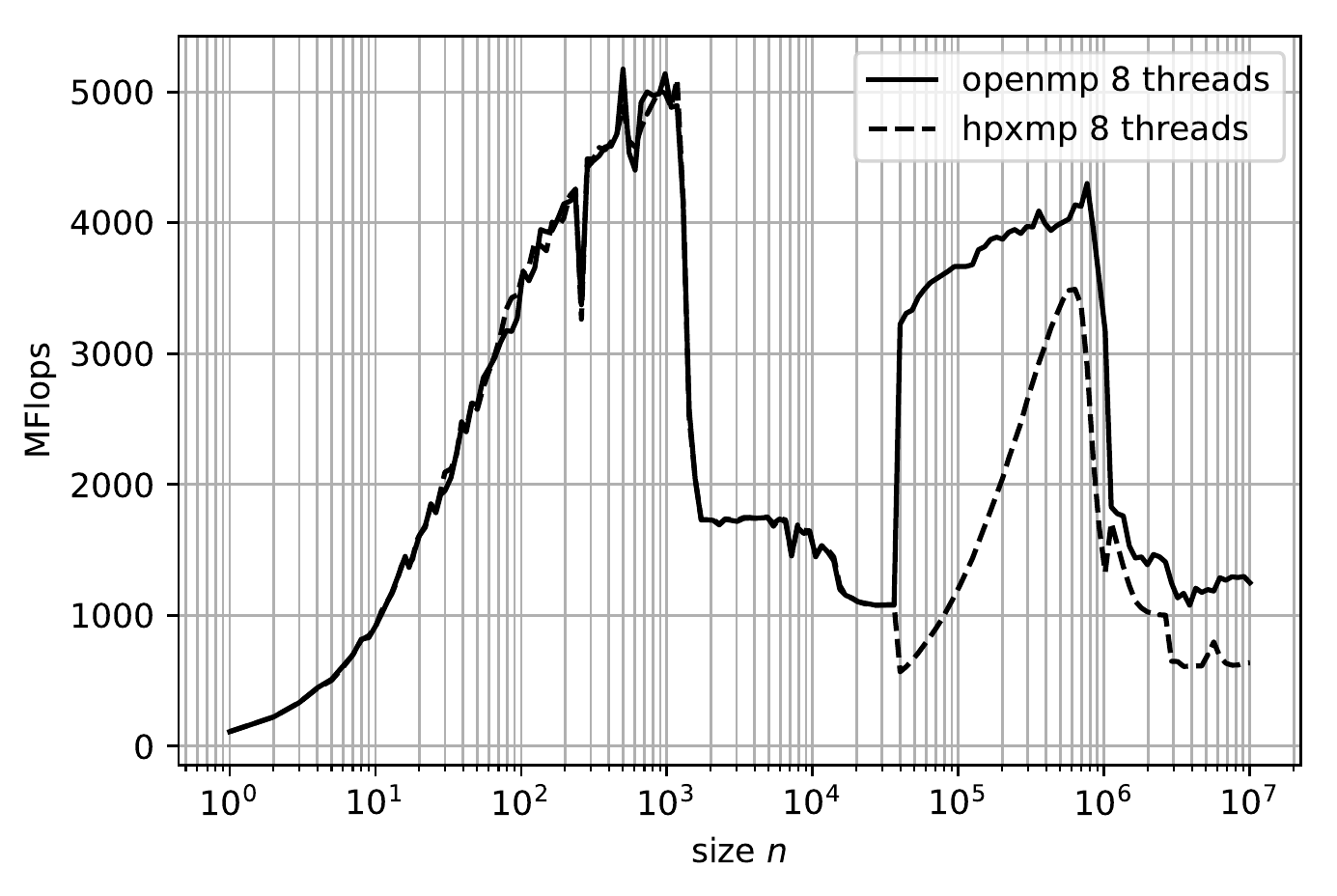}
		\caption{}
		\label{fig:ratio,dvecdvecadd_8}
	\end{subfigure}
	\begin{subfigure}[b]{0.3\textwidth}
		\includegraphics[width=\linewidth]{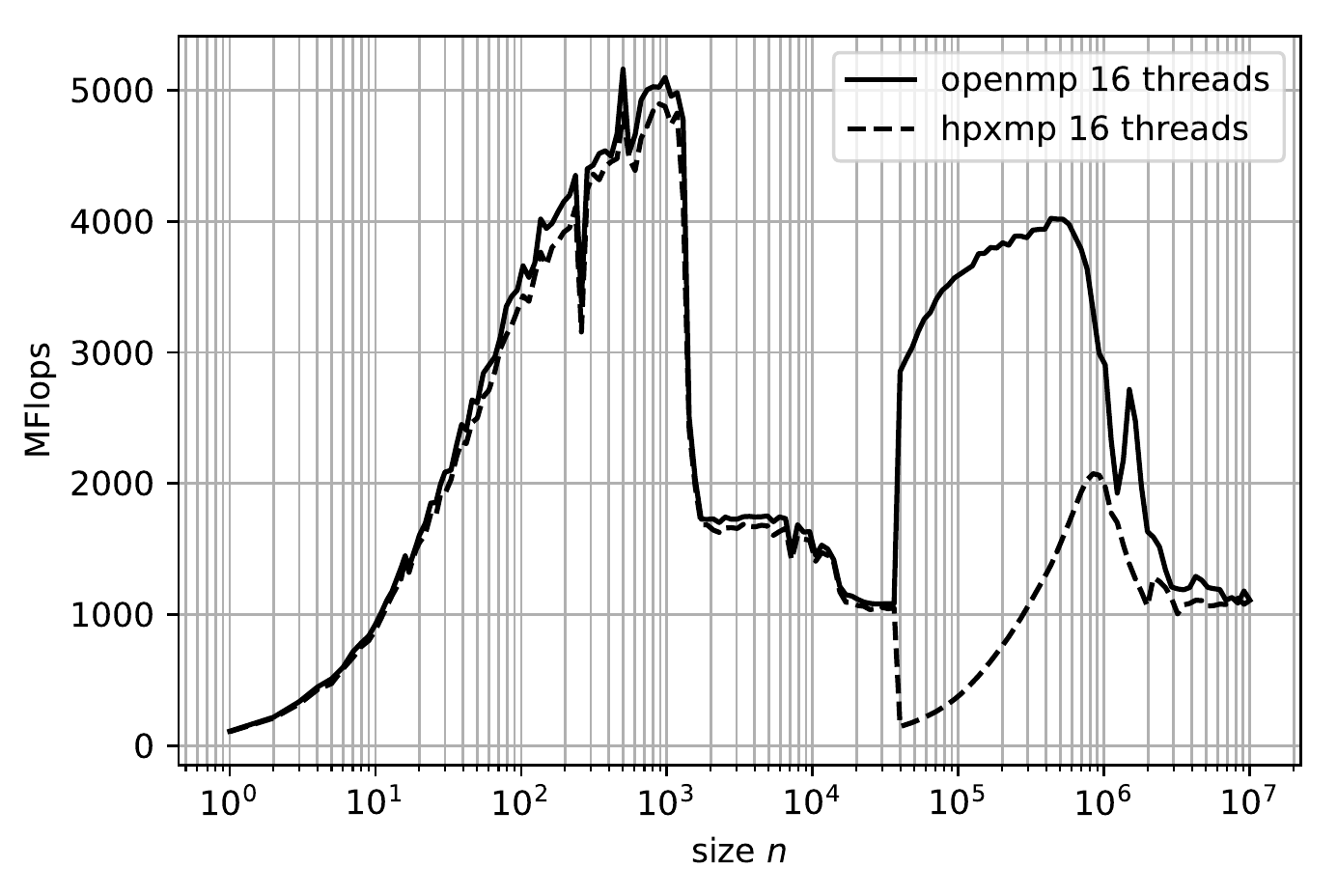}
		\caption{}
		\label{fig:ratio,dvecdvecadd_16}
	\end{subfigure}
	\caption{Scaling plots for dvecdvecadd Benchmarks for different number of threads: (a) $4$, (b) $8$, and (c) $16$}
	\label{fig:scale,dvecdvecadd}
\end{figure*}

\begin{figure*}[p]
	\centering
	\begin{subfigure}[b]{0.3\textwidth}
		\includegraphics[width=\linewidth]{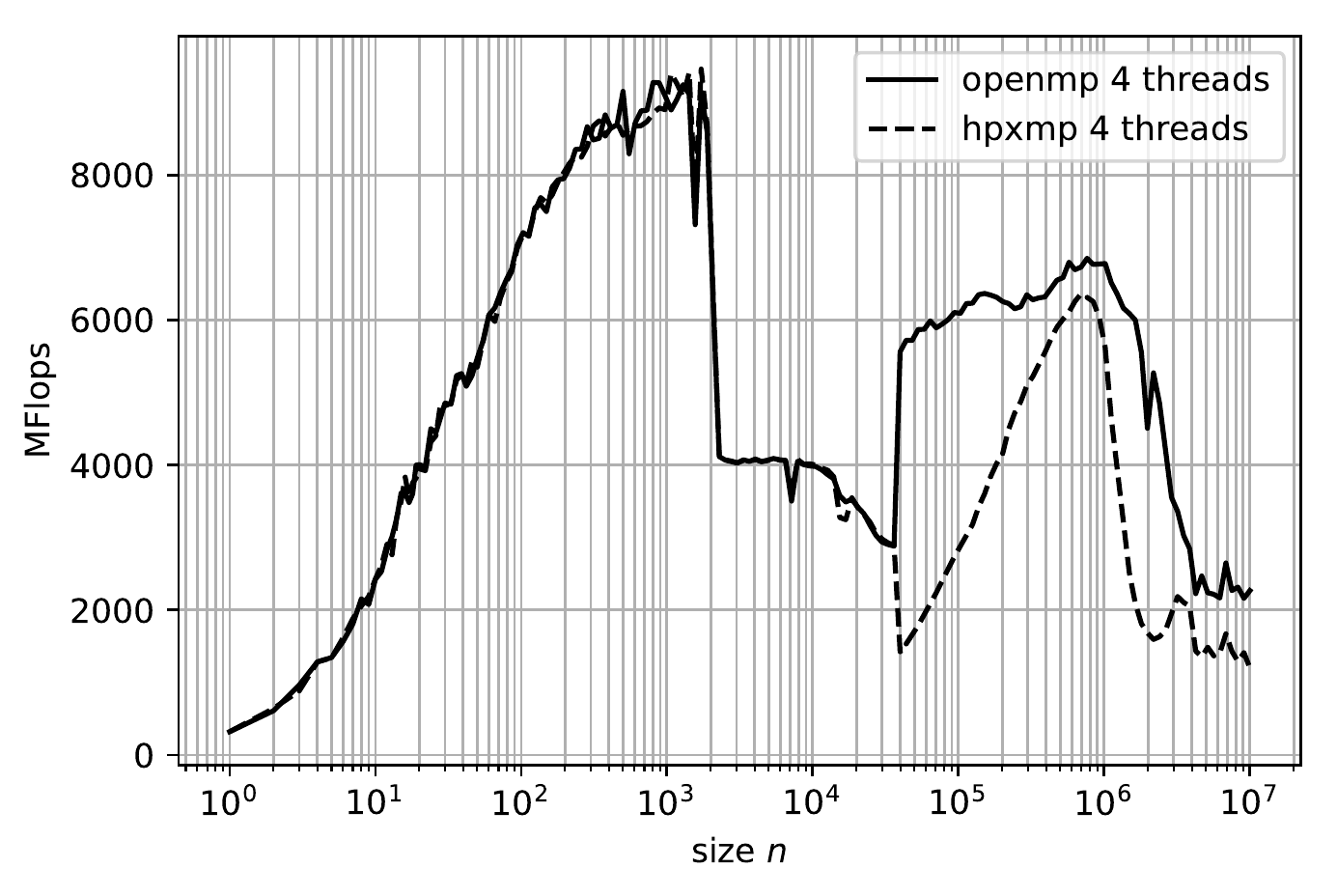}
		\caption{}
		\label{fig:ratio,daxpy_4}
	\end{subfigure}
	\begin{subfigure}[b]{0.3\textwidth}
		\includegraphics[width=\linewidth]{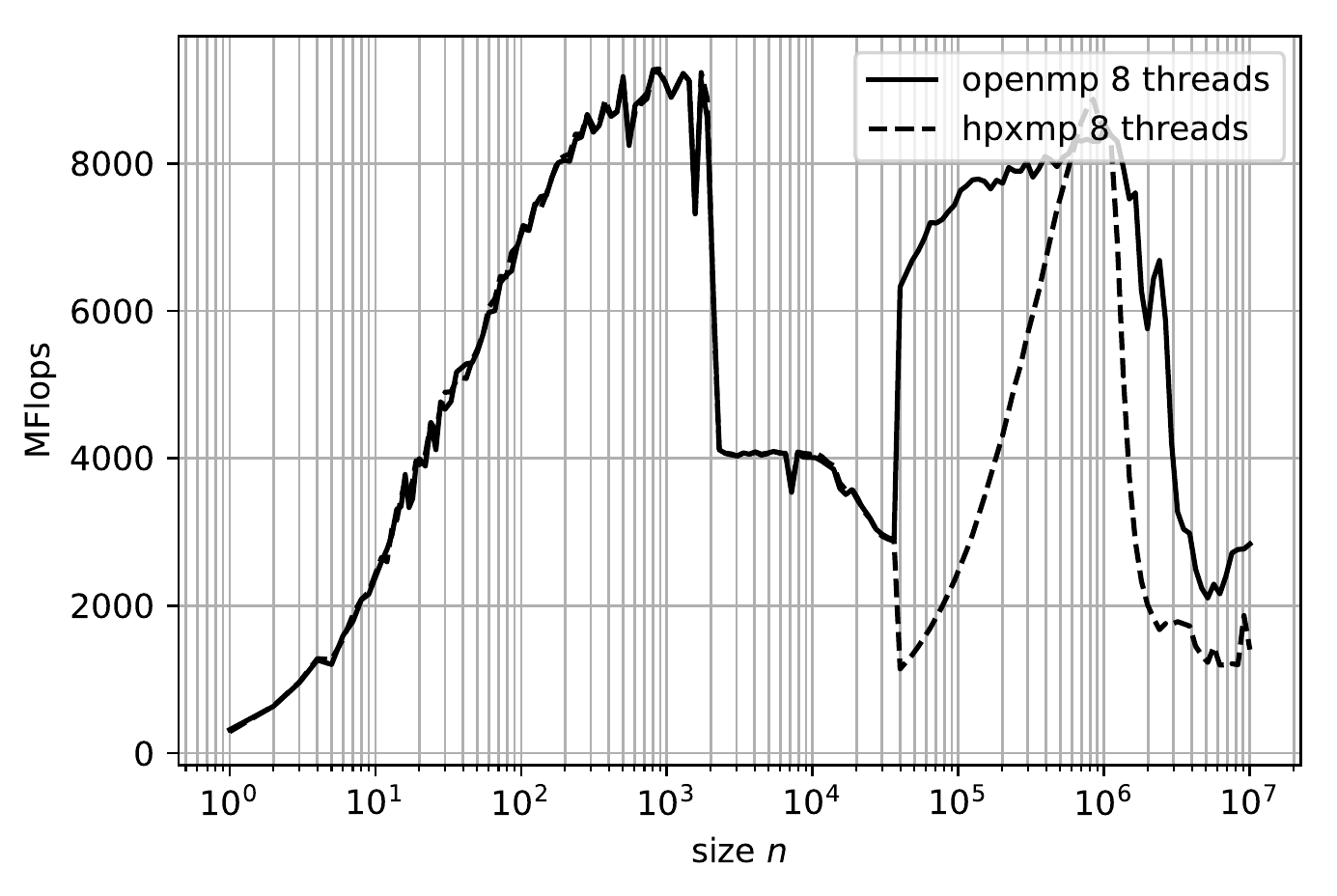}
		\caption{}
		\label{fig:ratio,daxpy_8}
	\end{subfigure}
	\begin{subfigure}[b]{0.3\textwidth}
		\includegraphics[width=\linewidth]{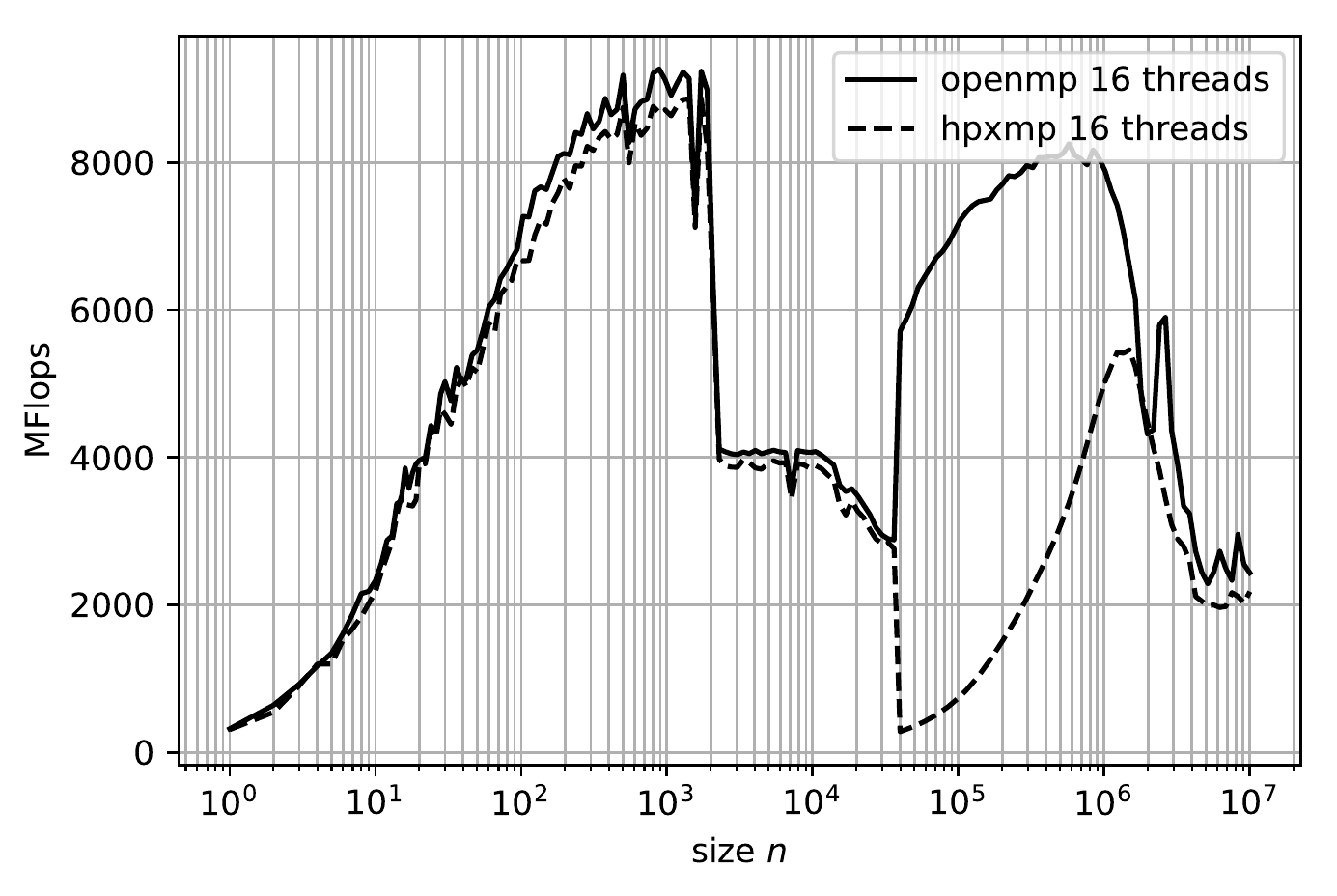}
		\caption{}
		\label{fig:ratio,daxpy_16}
	\end{subfigure}
	\caption{Scaling plots for daxpy Benchmarks for different number of threads: (a) $4$, (b) $8$, and (c) $16$}
	\label{fig:scale,daxpy}
\end{figure*}

\begin{figure*}[p]
	\centering
	\begin{subfigure}[b]{0.3\textwidth}
		\includegraphics[width=\linewidth]{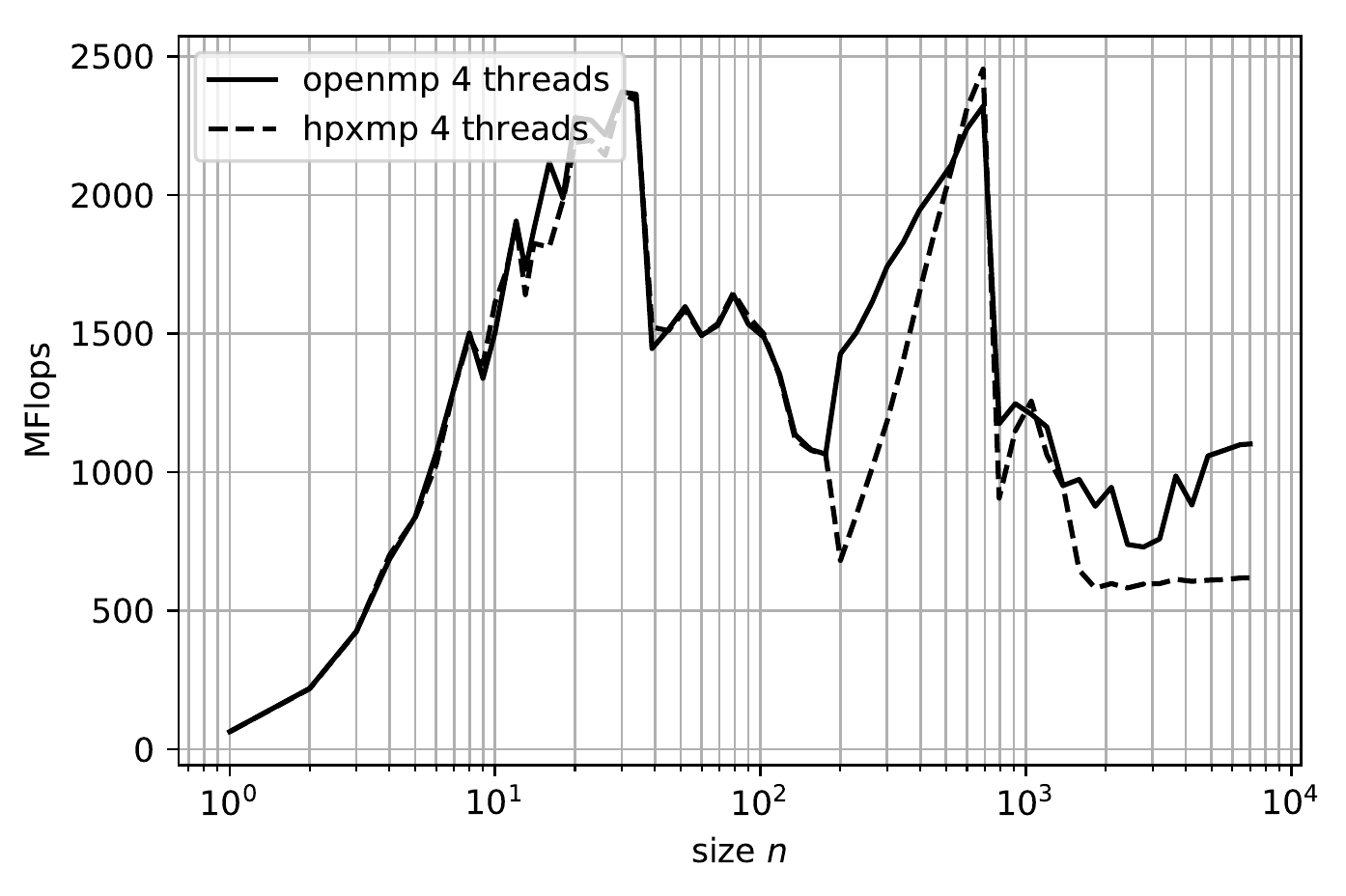}
		\caption{}
		\label{fig:ratio,dmatdmatadd_4}
	\end{subfigure}
	\begin{subfigure}[b]{0.3\textwidth}
		\includegraphics[width=\linewidth]{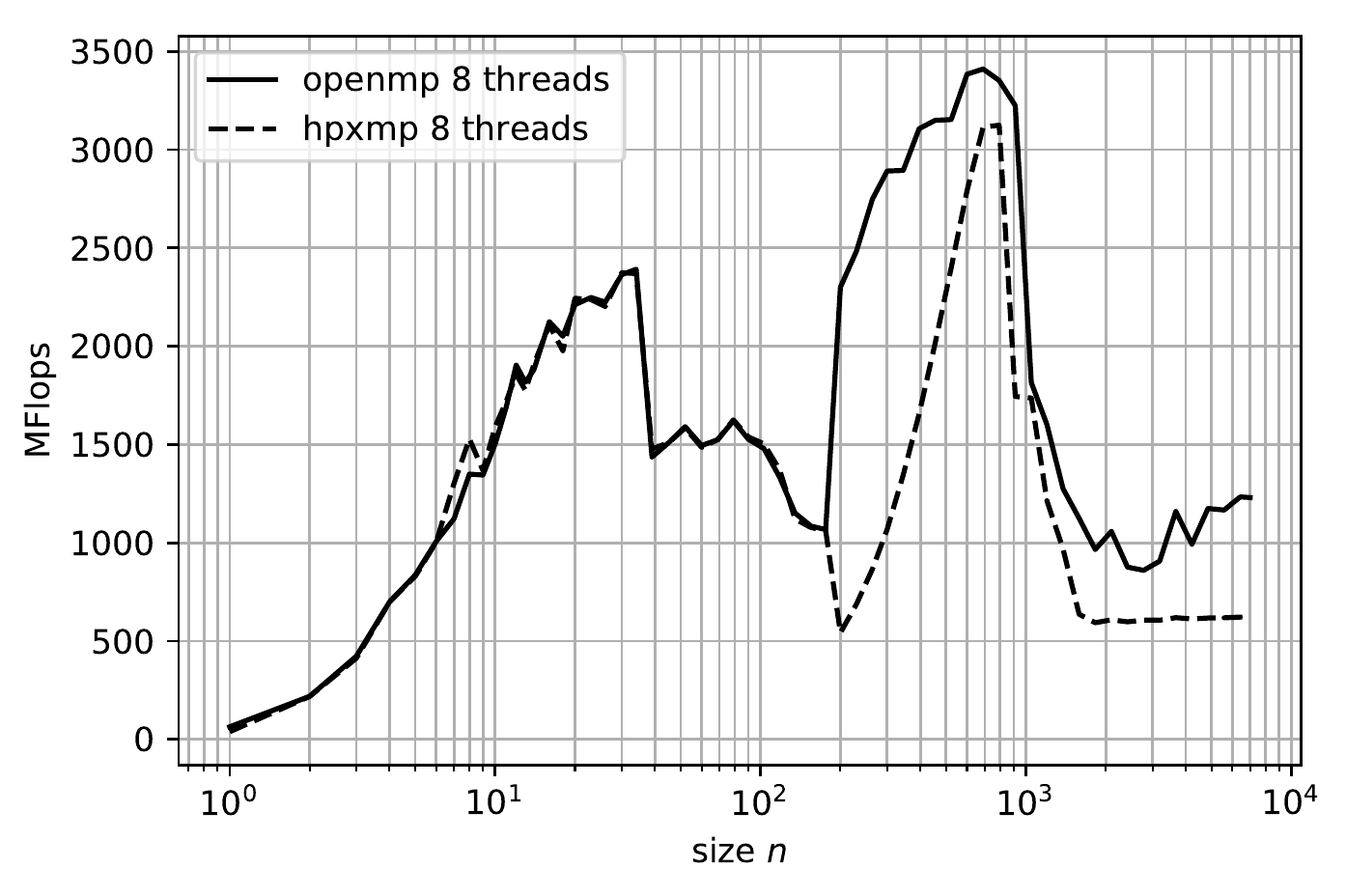}
		\caption{}
		\label{fig:ratio,dmatdmatadd_8}
	\end{subfigure}
	\begin{subfigure}[b]{0.3\textwidth}
		\includegraphics[width=\linewidth]{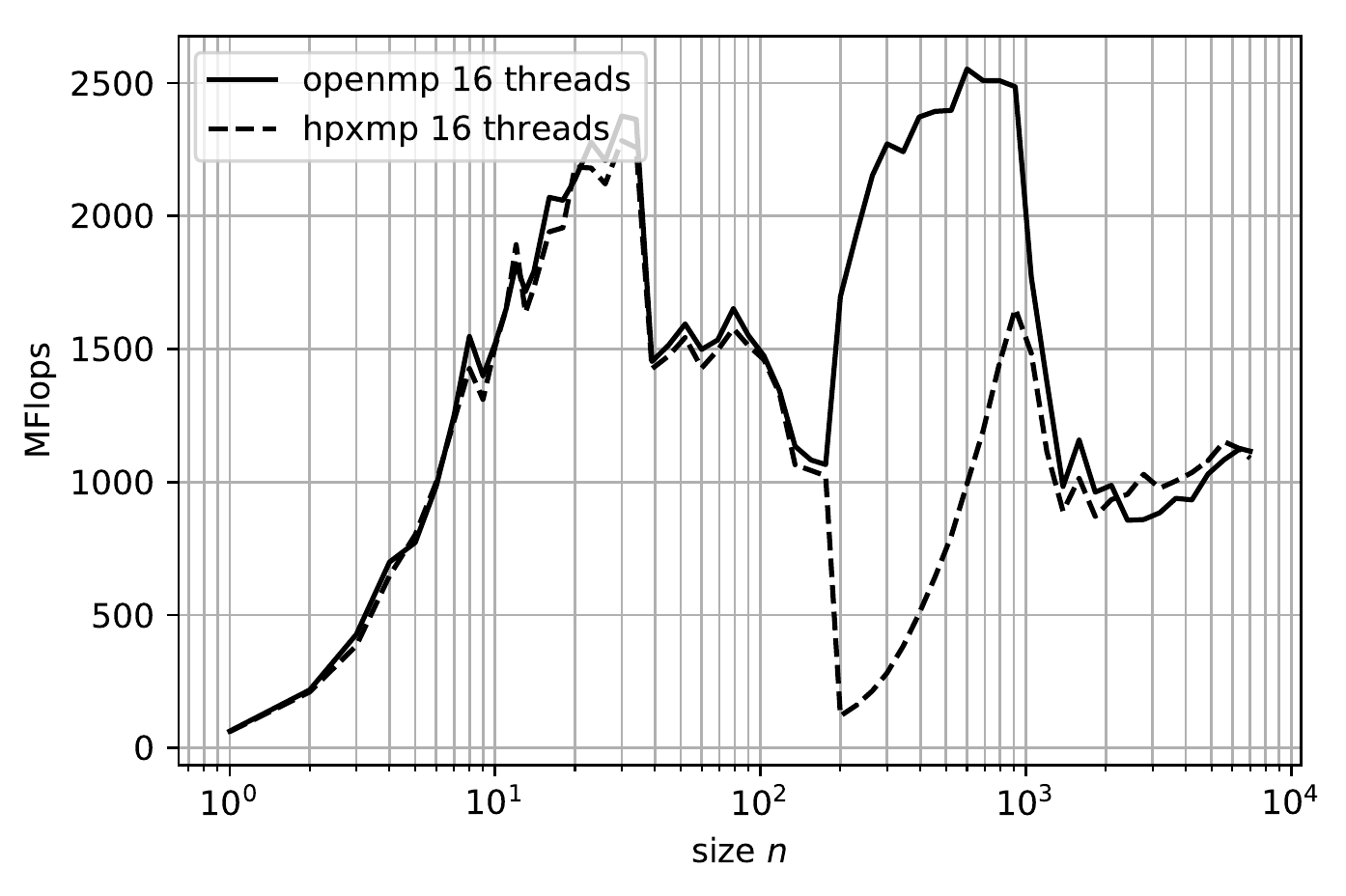}
		\caption{}
		\label{fig:ratio,dmatdmatadd_16}
	\end{subfigure}
	\caption{Scaling plots for dmatdmatadd Benchmarks for different number of threads: (a) $4$, (b) $8$, and (c) $16$}
	\label{fig:scale,dmatdmatadd}
\end{figure*}

\begin{figure*}[p]
	\centering
	\begin{subfigure}[b]{0.3\textwidth}
		\includegraphics[width=\linewidth]{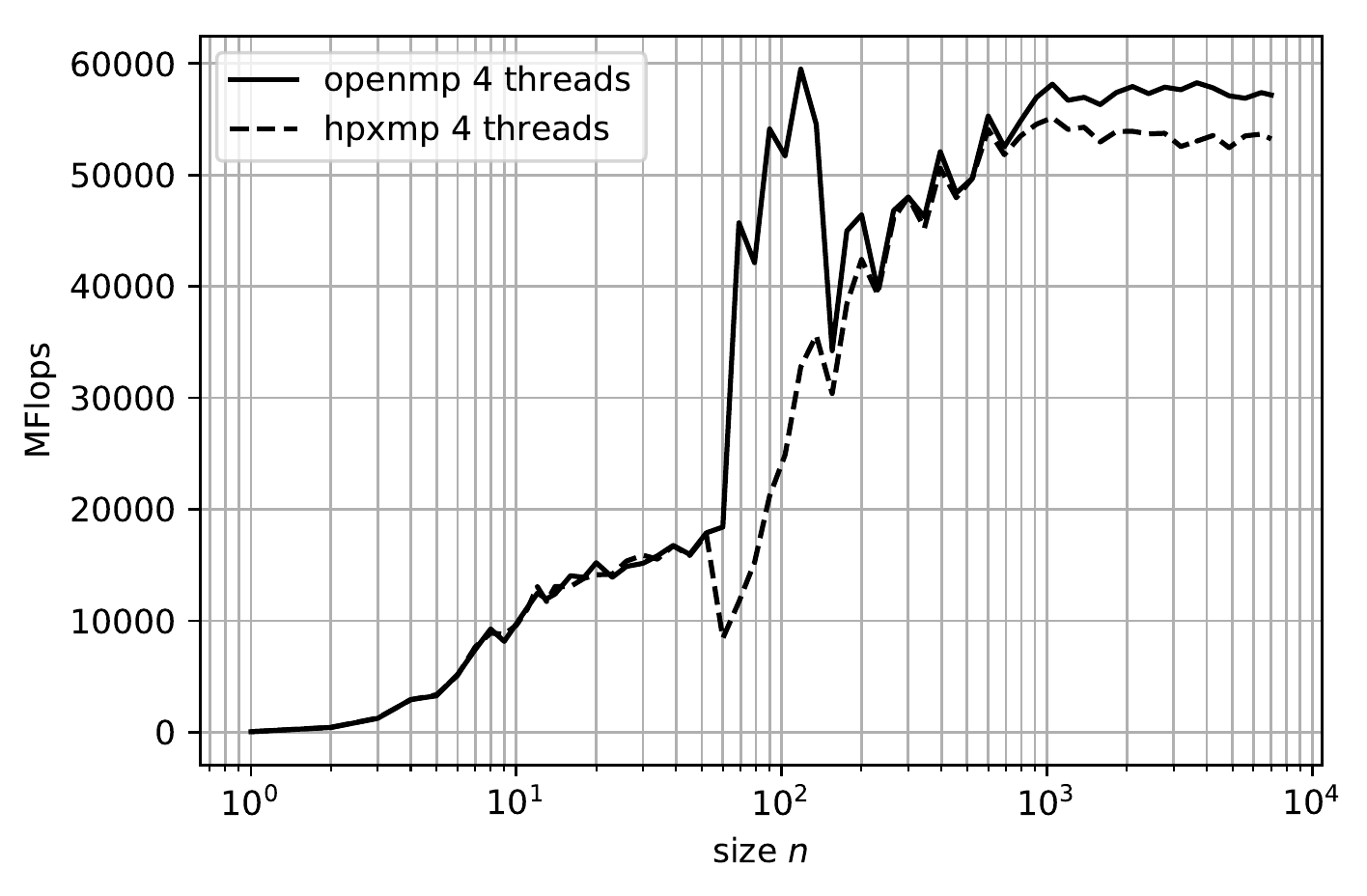}
		\caption{}
		\label{fig:ratio,dmatdmatmult_4}
	\end{subfigure}
	\begin{subfigure}[b]{0.3\textwidth}
		\includegraphics[width=\linewidth]{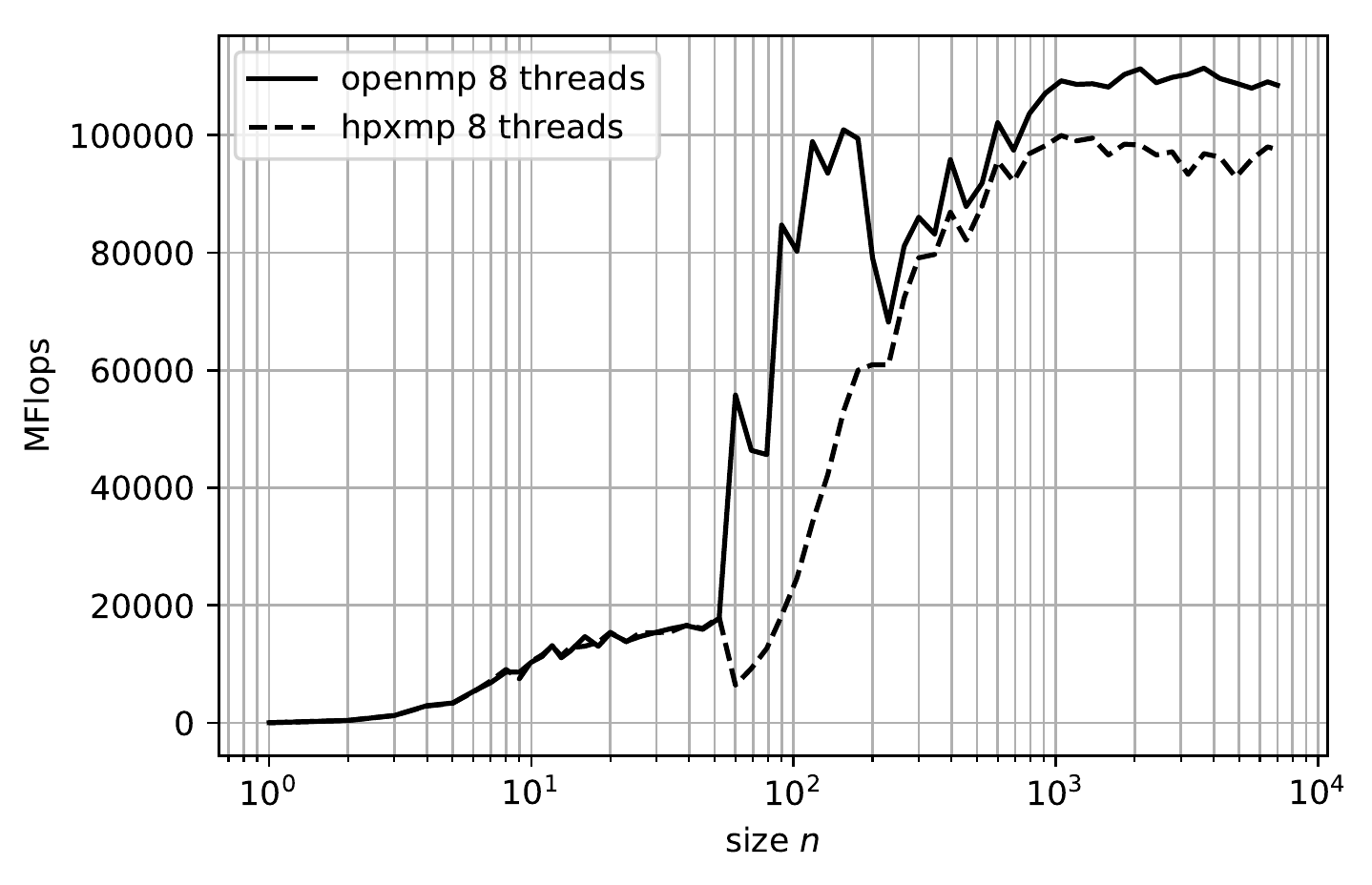}
		\caption{}
		\label{fig:ratio,dmatdmatmult_8}
	\end{subfigure}
	\begin{subfigure}[b]{0.3\textwidth}
		\includegraphics[width=\linewidth]{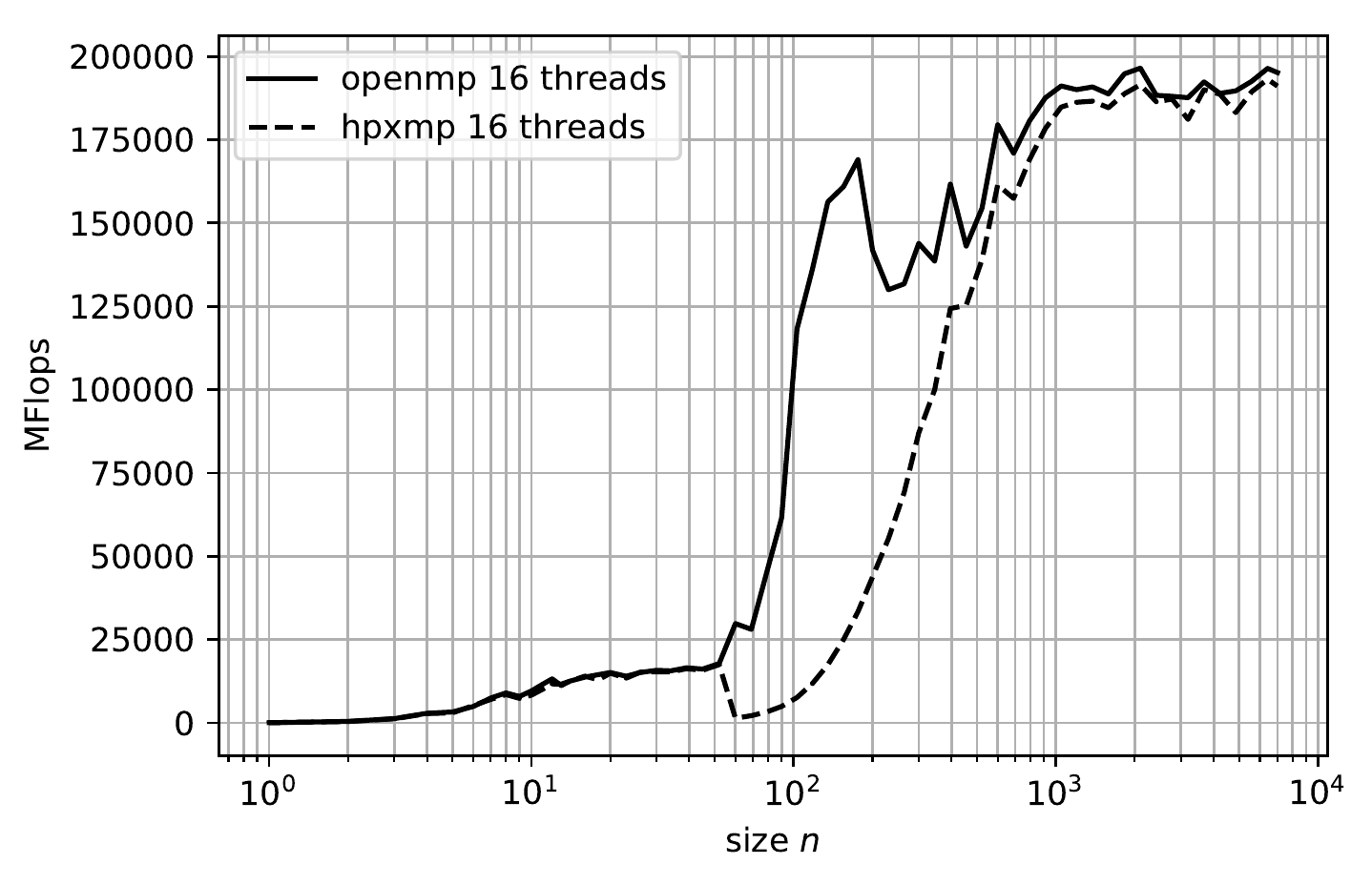}
		\caption{}
		\label{fig:ratio,dmatdmatmult_16}
	\end{subfigure}
	\caption{Scaling plots for dmatdmatmult Benchmarks for different number of threads: (a) $4$, (b) $8$, and (c) $16$}
	\label{fig:scale,dmatdmatmult}
\end{figure*}

\section{Conclusion and Outlook}
\label{sec:con}

This paper presents the design and architecture of an OpenMP runtime library built on top of HPX that supports most of the OpenMP V$3$ specification. We have demonstrated its full functionality by running various linear algebra benchmarks of the Blaze C++ library that for the tested functionalities relies on OpenMP for its parallelization needs. By replacing the compiler-supplied OpenMP runtime with our own, we were able to compare the performance of the two implementations. In general, our implementation is not able to reach the same performance compared to the native OpenMP solution yet. This is in part caused by Blaze being optimized for the compiler-supplied OpenMP implementations. We will work on optimizing the performance of hpxMP in the future. We have however demonstrated the viability of our solution for providing a smooth migration for applications that either directly or indirectly depend on OpenMP but have to be extended to benefit from a more general task based programming model.

\section*{Acknowledgment}
We thank Jeremy Kemp for providing the initial implementation of hpxMP\footnote{\url{https://github.com/kempj/hpxMP}} which was extended by the authors. The work on hpxMP is funded by the National Science Foundation (award 1737785).

\appendix

\section{Source code}
The source code of hpxMP is available on github\footnote{\url{https://github.com/STEllAR-GROUP/hpxMP}} released under the BSL $1$.$0$.

\bibliographystyle{unsrt} 
\bibliography{sample-base,../../bibfiles/hpx} 

\begin{thebibliography}{10}

\bibitem{openmp_spec}
{OpenMP Consortium}.
\newblock {OpenMP Specification Version 5.0}.
\newblock Technical report, OpenMP Consortium, 2018.
\newblock
  \url{https://www.openmp.org/wp-content/uploads/OpenMP-API-Specification-5.0.pdf}.

\bibitem{10.1109/MC.2016.232}
David~A. Bader.
\newblock Evolving mpi+x toward exascale.
\newblock {\em Computer}, 49(8):10, 2016.

\bibitem{heller2017hpx}
Thomas Heller, Patrick Diehl, Zachary Byerly, John Biddiscombe, and Hartmut
  Kaiser.
\newblock Hpx--an open source c++ standard library for parallelism and
  concurrency.
\newblock 2017.

\bibitem{cxx11_standard}
{C++ Standards Committee}.
\newblock {ISO/IEC 14882:2011, Standard for Programming Language C++ (C++11)}.
\newblock Technical report, ISO/IEC JTC1/SC22/WG21 (the C++ Standards
  Committee), 2011.
\newblock \url{https://wg21.link/N3337}, last publicly available draft.

\bibitem{cxx17_standard}
{C++ Standards Committee}.
\newblock {ISO/IEC DIS 14882, Standard for Programming Language C++ (C++17)}.
\newblock Technical report, ISO/IEC JTC1/SC22/WG21 (the C++ Standards
  Committee), 2017.
\newblock \url{https://wg21.link/N4659}, last publicly available draft.

\bibitem{heller_gb}
Thomas Heller, Bryce Lelbach, Kevin Huck, John Biddiscombe, Patricia Grubel,
  Alice Koniges, Matthias Kretz, Dominic Marcello, David Pfander, Adrian Serio,
  Juhan Frank, Geoffrey Clayton, Dirk Pflüger, David Eder, and Hartmut Kaiser.
\newblock Harnessing billions of tasks for a scalable portable hydrodynamic
  simulation of the merger of two stars.
\newblock {\em International Journal of High Performance Computing Applications
  (IJHPCA)}, 2018.

\bibitem{eigenweb}
Ga\"{e}l Guennebaud, Beno\^{i}t Jacob, et~al.
\newblock Eigen v3.
\newblock http://eigen.tuxfamily.org, 2010.

\bibitem{rupp2016viennacl}
Karl Rupp, Philippe Tillet, Florian Rudolf, Josef Weinbub, Andreas Morhammer,
  Tibor Grasser, Ansgar J\"ungel, and Siegfried Selberherr.
\newblock Viennacl---linear algebra library for multi-and many-core
  architectures.
\newblock {\em SIAM Journal on Scientific Computing}, 38(5):S412--S439, 2016.

\bibitem{sanderson2016armadillo}
Conrad Sanderson and Ryan Curtin.
\newblock Armadillo: a template-based c++ library for linear algebra.
\newblock {\em Journal of Open Source Software}, 2016.

\bibitem{iglberger2012high}
Klaus Iglberger, Georg Hager, Jan Treibig, and Ulrich R{\"u}de.
\newblock High performance smart expression template math libraries.
\newblock In {\em High Performance Computing and Simulation (HPCS), 2012
  International Conference on}, pages 367--373. IEEE, 2012.

\bibitem{galassi2002gnu}
Mark Galassi, Jim Davies, James Theiler, Brian Gough, Gerard Jungman, Patrick
  Alken, Michael Booth, and Fabrice Rossi.
\newblock Gnu scientific library.
\newblock {\em Network Theory Ltd}, 3, 2002.

\bibitem{wang2013augem}
Qian Wang, Xianyi Zhang, Yunquan Zhang, and Qing Yi.
\newblock Augem: automatically generate high performance dense linear algebra
  kernels on x86 cpus.
\newblock In {\em High Performance Computing, Networking, Storage and Analysis
  (SC), 2013 International Conference for}, pages 1--12. IEEE, 2013.

\bibitem{laug}
E.~Anderson, Z.~Bai, C.~Bischof, S.~Blackford, J.~Demmel, J.~Dongarra,
  J.~Du~Croz, A.~Greenbaum, S.~Hammarling, A.~McKenney, and D.~Sorensen.
\newblock {\em {LAPACK} Users' Guide}.
\newblock Society for Industrial and Applied Mathematics, Philadelphia, PA,
  third edition, 1999.

\bibitem{blackford2002updated}
L~Susan Blackford, Antoine Petitet, Roldan Pozo, Karin Remington, R~Clint
  Whaley, James Demmel, Jack Dongarra, Iain Duff, Sven Hammarling, Greg Henry,
  et~al.
\newblock An updated set of basic linear algebra subprograms (blas).
\newblock {\em ACM Transactions on Mathematical Software}, 28(2):135--151,
  2002.

\bibitem{pthreads}
Robert~A. Alfieri.
\newblock An efficient kernel-based implementation of {POSIX} threads.
\newblock In {\em Proceedings of the USENIX Summer 1994 Technical Conference on
  USENIX Summer 1994 Technical Conference - Volume 1}, USTC'94, pages 5--5,
  Berkeley, CA, USA, 1994. USENIX Association.

\bibitem{inteltbb}
Intel.
\newblock {Intel Thread Building Blocks}, 2019.
\newblock http://www.threadingbuildingblocks.org.

\bibitem{microsoftppl}
Microsoft.
\newblock {Microsoft Parallel Pattern Library}, 2010.
\newblock http://msdn.microsoft.com/en-us/library/dd492418.aspx.

\bibitem{chamberlain07parallelprogrammability}
Bradford~L. Chamberlain, David Callahan, and Hans~P. Zima.
\newblock {Parallel Programmability and the Chapel Language}.
\newblock {\em International Journal of High Performance Computing Applications
  (IJHPCA)}, 21(3):291--312, 2007.
\newblock \url{https://dx.doi.org/10.1177/1094342007078442}.

\bibitem{cilk++}
Charles~E. Leiserson.
\newblock {The Cilk++ concurrency platform}.
\newblock In {\em DAC '09: Proceedings of the 46th Annual Design Automation
  Conference}, pages 522--527, New York, NY, USA, 2009. ACM.

\bibitem{CarterEdwards20143202}
H.~Carter Edwards, Christian~R. Trott, and Daniel Sunderland.
\newblock Kokkos: Enabling manycore performance portability through polymorphic
  memory access patterns.
\newblock {\em Journal of Parallel and Distributed Computing}, 74(12):3202 --
  3216, 2014.
\newblock Domain-Specific Languages and High-Level Frameworks for
  High-Performance Computing.

\bibitem{openmp1}
Leonardo Dagum and Ramesh Menon.
\newblock Open{MP}: An industry-standard api for shared-memory programming.
\newblock {\em IEEE Computational Science and Engineering}, 5(1):46--55, 1998.

\bibitem{ppl_link}
{PPL}.
\newblock {PPL - Parallel Programming Laboratory}, 2011.
\newblock {http://charm.cs.uiuc.edu/}.

\bibitem{barrett2015toward}
Richard~F Barrett, Dylan~T Stark, Courtenay~T Vaughan, Ryan~E Grant, Stephen~L
  Olivier, and Kevin~T Pedretti.
\newblock Toward an evolutionary task parallel integrated mpi+ x programming
  model.
\newblock In {\em Proceedings of the Sixth International Workshop on
  Programming Models and Applications for Multicores and Manycores}, pages
  30--39. ACM, 2015.

\bibitem{khatami2016massively}
Zahra Khatami, Hartmut Kaiser, Patricia Grubel, Adrian Serio, and J~Ramanujam.
\newblock A massively parallel distributed n-body application implemented with
  hpx.
\newblock In {\em 2016 7th Workshop on Latest Advances in Scalable Algorithms
  for Large-Scale Systems (ScalA)}, pages 57--64. IEEE, 2016.

\bibitem{copik2017using}
Marcin Copik and Hartmut Kaiser.
\newblock Using sycl as an implementation framework for hpx. compute.
\newblock In {\em Proceedings of the 5th International Workshop on OpenCL},
  page~30. ACM, 2017.

\bibitem{wagle2018methodology}
Bibek Wagle, Samuel Kellar, Adrian Serio, and Hartmut Kaiser.
\newblock Methodology for adaptive active message coalescing in task based
  runtime systems.
\newblock In {\em 2018 IEEE International Parallel and Distributed Processing
  Symposium Workshops (IPDPSW)}, pages 1133--1140. IEEE, 2018.

\bibitem{khatami2017redesigning}
Zahra Khatami, Hartmut Kaiser, and J~Ramanujam.
\newblock Redesigning op2 compiler to use hpx runtime asynchronous techniques.
\newblock {\em arXiv preprint arXiv:1703.09264}, 2017.

\bibitem{biddiscombezero}
John Biddiscombe, Anton Bikineev, Thomas Heller, and Hartmut Kaiser.
\newblock Zero copy serialization using rma in the hpx distributed task-based
  runtime.
\newblock 2017.

\bibitem{grubel2015performance}
Patricia Grubel, Hartmut Kaiser, Jeanine Cook, and Adrian Serio.
\newblock The performance implication of task size for applications on the hpx
  runtime system.
\newblock In {\em Cluster Computing (CLUSTER), 2015 IEEE International
  Conference on}, pages 682--689. IEEE, 2015.

\bibitem{openmpintro}
Tim Mattson.
\newblock A ''hands-on'' introduction to openmp, 2013.

\bibitem{openmptr}
Bronis~R. de~Supinski Michael~Klemm.
\newblock Openmp technical report 6:version 5.0 preview 2.
\newblock Technical report, OpenMP Architecture Review Board, 2017.

\end{thebibliography}

\end{document}